\documentclass[usenames,dvipsnames]{aastex631}

  \renewcommand{\d} {\mathrm{d}}         
  
 \newcommand{\vct}[1]  {\ensuremath{\boldsymbol{#1}}}    

  \newcommand{\uv}[1]   {\hat{\vct{#1}}}  
  \newcommand {\vz}     {\vct{z}}
  \newcommand {\vY}     {\vct{Y}}   
  \newcommand {\divYpm} {$ \nabla_\ell \cdot \vY^\pm $}
  \newcommand{\epsDiss} {\epsilon_\text{diss}}
   
  \newcommand{\vKH} {von K\'arm\'an--Howarth } 

\graphicspath{ {images/} }
\usepackage{amsmath}
\usepackage{amssymb}
\usepackage{relsize}
\usepackage{mathtools}
\usepackage{hyperref}
\usepackage{float}
\usepackage{tabularx}
\usepackage{color}
\usepackage{tikz}
\usetikzlibrary{arrows, petri, topaths}
\usepackage{hyperref}
\usepackage{colortbl}
\usepackage{fancyhdr}
\usepackage{physics}
\usepackage{braket}

\usepackage{natbib}
\usepackage{verbatim} 

\graphicspath{{./}{figures/}}

\begin{document}

\title{Strategies for determining the cascade rate in MHD turbulence: isotropy, anisotropy, and spacecraft sampling}

\correspondingauthor{William H. Matthaeus}
\email{whm@udel.edu}


\author{Yanwen Wang}
\affiliation{Department of Physics and Astronomy, University of Delaware, USA}


\author[0000-0002-7174-6948]{Rohit Chhiber}
\affiliation{Department of Physics and Astronomy, University of Delaware, USA}
\affiliation{
Heliophysics Science Division, NASA Goddard Space Flight Center, USA}

\author[0000-0003-2160-7066]{Subash Adhikari}
\affiliation{Department of Physics and Astronomy, University of Delaware, USA}

\author[0000-0003-2965-7906]{Yan Yang}
\affiliation{Department of Physics and Astronomy, University of Delaware, USA}

\author[0000-0002-6962-0959]{Riddhi Bandyopadhyay}
\affiliation{Department of Astrophysical Sciences, Princeton University, USA}

\author{Michael A. Shay}
\affiliation{Department of Physics and Astronomy, University of Delaware, USA}

\author[0000-0002-2814-7288]{Sean Oughton}
\affiliation{Department of Mathematics, University of Waikato, Hamilton, New Zealand}

\author[0000-0001-7224-6024]{William H. Matthaeus}
\affiliation{Department of Physics and Astronomy, University of Delaware, USA}

\author[0000-0002-7341-2992]{Manuel E. Cuesta}
\affiliation{Department of Physics and Astronomy, University of Delaware, USA}



\begin{abstract}

``Exact'' laws for evaluating cascade rates, tracing back to the Kolmogorov ``4/5'' law, have been extended to many systems of interest including magnetohydrodynamics (MHD),
and compressible flows of the magnetofluid and ordinary fluid types. It is understood that 
implementations may be limited by the quantity of 
available data and by the lack of turbulence symmetry.  
Assessment of the accuracy and feasibility of such ``third-order'' (or Yaglom) relations is most effectively accomplished by examining the \vKH equation in increment form, a framework from which the third-order laws are derived as asymptotic approximations. Using this approach, we examine the context of third-order laws for incompressible MHD in some detail. 
The simplest versions rely on the assumption of isotropy
and the presence of a well-defined inertial range, 
while related procedures generalize the same idea to arbitrary rotational symmetries. 
Conditions for obtaining correct and accurate values of the dissipation rate from these laws based on several sampling 
and fitting  strategies are investigated using results from simulations. 
The questions we address
are of particular relevance to sampling of solar wind turbulence by one or more spacecraft.

\end{abstract}



\section{Introduction}\label{sec:intro}

\nocite{Kol41a,Kol41c} 

Direct measures of cascade rates in turbulent systems often 
employ theoretical 
formulations related to Kolmogorov's ``4/5'' law 
    \citep{Kol41c,Frisch}
and its variants, 
in which the inertial range cascade rate is related to 
a signed third-order structure function. 
This so-called ``exact'' law is derived from the fluid equations 
without appeal to dimensional analysis, assumptions about scaling behavior, 
or any ansatz concerning time scales; however, this law does require time stationarity, spatial homogeneity,
the existence of an inertial range, and 
a finite dissipation rate. 
The original formulation for isotropic incompressible hydrodynamics
has been extended to magnetohydrodynamics (MHD) 
\citep{PolitanoPouquet98-vKH,PolitanoPouquet98-grl}
and related models.
The MHD version is frequently applied to in situ observations of plasma turbulence in the solar wind 
\citep{Sorriso-ValvoEA07,MacBrideEA08,BandyopadhyayEA20-PSP3rd}
to obtain cascade rates that inform theories of heating and
acceleration of the solar wind \citep{OsmanEA11-3rd},
providing ground truth for related approximations 
in space physics \citep{VasquezEA07-cascade}.
Frequently a 
major issue in these applications is the use of 
formulations derived assuming 
isotropy in turbulence that is actually anisotropic \citep{VerdiniEA15}, 
this being the typical case 
for solar wind and magnetosheath turbulence. 
Usually this potential inconsistency 
is disregarded in favor of extensive averaging, whenever possible.
Another more practical 
limitation is the challenging
requirement 
of a sufficient volume of data
\citep{PodestaEA09-3rd}, 
a kinematic and statistical issue further 
complicated by potential sensitivity to the tails of
the probability distribution of the fluctuations
\citep{Dudok04}.
Taking these 
challenges into account, 
we note that the ability to extract cascade rates from observational data is of increasing importance 
due to the centrality of fundamental questions relating 
to heating and dissipation in space and astrophysical plasmas \citep[e.g.,][]{KiyaniEA15}.
Therefore in the present study we revisit several related issues 
that are pertinent to 
the evaluation   
of third-order laws 
using single-point or multi-point measurements.
We re-examine the issue of averaging by focusing on 
conditions for obtaining accurate results in both isotropic and anisotropic turbulence. The strategies we examine are 
implemented  using data from  three-dimensional (3D) MHD 
turbulence simulations.  
A motivation
for this  approach  is that for such cases we have an unambiguous determination of the underlying turbulence 
symmetry as well as a straightforward method to 
quantify the absolute dissipation rate. 

The remainder of the paper is structured as follows.
In Section~\ref{sec:background} we review relevant theoretical and observational studies that set the stage for
the questions we address. 
Section~\ref{sec:sims} describes in detail 
the simulations used for the present study.
Section~\ref{sec:results_1Dlaw} contains the results for the one-dimensional (1D) form of the third-order law using  measurements for the isotropic and anisotropic cases. 
Section~\ref{sec:results_3Dlaw} delves into the direction-averaged 1D form of the third-order law in the inertial range
and shows the effect of all terms in the \vKH equation.
Section~\ref{sec:spacecraft} gives an example of using the 1D form third-order law in a single spacecraft sample, and discusses the relative accuracy of this strategy to estimate the energy dissipation rate in observational measurements.  
Section~\ref{sec:conc} provides a summary of the results, and examines the relationship between the strategies used in this study and their potential applications to multi-point observations via a constellation of spacecraft.

\section{Background}\label{sec:background}
\subsection{Theory}\label{subsec:theory}

We start from the 3D incompressible MHD equations \citep[e.g.,][]{Biskamp-turb}
\begin{equation}
    \begin{split}
    \frac{\partial \vct{v}}{\partial t} + \vct{v} \cdot \grad \vct{v} &= -\grad P + \vct{B} \cdot \grad \vct{B} + \nu \laplacian \vct{v} , \\
    \frac{\partial \vct{B}}{\partial t} + \vct{v} \cdot \grad \vct{B} &= \vct{B} \cdot \grad \vct{v} + \mu \laplacian \vct{B},
    \label{eq:vb}
    \end{split}
\end{equation}
where $\vct{v}$ and $\vct{B}$ represent the local velocity and magnetic field (the latter in Alfv\'en speed units with $ \vct{B}/\sqrt{4\pi\rho} \to \vct{B}$ and uniform mass density $\rho$), 
$P$ is the total (thermal plus magnetic) pressure, 
$\nu$ is the kinematic viscosity, and $\mu$ is the resistivity. 
Note that $\vct{B} = \vct{B}_0 + \vct{b}$ where $\vct{B}_0$ is the global mean field and $\vct{b}$ is the fluctuating field. 
As is well known, one may work with the Els\"asser variables \citep{Elsasser50-pr},
  $\vz^{\pm}(\vct{r}) = \vct{v}(\vct{r}) \pm \vct{b}(\vct{r})$,
instead of $\vct{v}$ and $\vct{b}$.  For situations where 
    $ \nu = \mu $,
the incompressible MHD equations are then rewritten as:
\begin{equation}
\frac{\partial}{\partial t} \vz^{\pm} = -(\vz^{\mp} \cdot \grad)\vz^{\pm} - \grad P + \nu \laplacian \vz^{\pm} \pm (\vct{B}_0 \cdot \grad) \vz^{\pm} .
 \label{eq:zpm}
\end{equation}

By taking the difference of the equation at $\vz^{\pm}(\vct{r})$ and that at $\vz^{\pm}(\vct{r} +  \vct{\ell} )$, and assuming homogeneity and incompressibility, the pressure term and the term containing $\vct{B}_0$ vanish. 
Let us define
\begin{equation}
    \delta \vz^{\pm}(\vct{r}, \vct{\ell}) = \vz^{\pm}(\vct{r} + \vct{\ell}) - \vz^{\pm}(\vct{r})
  \label{eq:incr-defn}
\end{equation}
as the \emph{increment} of the Els\"asser variables. 
Taking now the dot product of $ \delta \vz^\pm $ with the equation for $\partial (\delta \vz^\pm) / \partial t$ and performing an ensemble average of the result yields the MHD
\vKH equation \citep{PolitanoPouquet98-vKH,PolitanoPouquet98-grl}:
\begin{equation}
\frac{\partial}{\partial t}\langle (\delta \vz^{\pm})^2 \rangle = -\grad_{\ell} \cdot \langle \delta \vz^{\mp} |\delta \vz^{\pm}|^2 \rangle + 2\nu \laplacian_{\ell} \langle (\delta \vz^{\pm})^2\rangle - 4\epsilon^{\pm}.
\label{eq:fullform3d}
\end{equation}
Recall that because of homogeneity taking either an ensemble average \citep{McComb} or a spatial average (denoted as $\langle \bullet \rangle$), means that the averaged increments $\langle \delta \vz^{\pm}( \vct{r}, \vct{\ell}) \rangle$ 
are only dependent on the lag $\vct{\ell}$, and similarly for other moments of the increments.
Here $\epsilon^{\pm} = \nu \langle [\grad  \vz^{\pm}(\vct{r})]^2\rangle$ are 
the mean dissipation rates associated with the Els\"asser energies
 $ \langle \vz^\pm \cdot \vz^\pm \rangle/2 $
(not the increments). 
Since these dissipation rates involve real-space gradients 
  (i.e., $\grad$ not $\grad_{\ell} $)
and are independent of lag, they are constant over all length scales. The total energy dissipation rate of the system, also lag independent, is  
\begin{equation}
   \epsDiss 
   = \frac{\epsilon^+ + \epsilon^-}{2}
   = 
   \nu \langle \omega^2 + J^2 \rangle,
 \label{eq:eps-diss}
\end{equation}
 where the second form is in terms of the mean-square vorticity 
$\langle \omega^2 \rangle = \langle (\curl \vct{v})^2\rangle$ and mean-square electric current density $\langle J^2 \rangle = \langle (\curl \vct{b})^2 \rangle $.

Eq.~\eqref{eq:fullform3d} is the fundamental equation of energy conservation on which all related results presented below will be based. The terms of Eq.~\eqref{eq:fullform3d} express four effects: 
time dependence, nonlinear transfer, 
dissipation (of the mean-square increments), 
and the 
exact dissipation 
rate of Els\"asser energies,
respectively from left to right.  
For convenience in referring to 
the first three of these terms, which are lag dependent, we designate
$\frac{\partial}{\partial t} \langle (\delta \vz^{\pm})^2 \rangle = T^{\pm}$,  
$\langle (\delta \vz^{\pm})^2 \rangle = G^{\pm}$,
and 
\begin{equation}
  \vct{Y}^\pm (\vct{\ell}) 
    = \langle \delta \vz^\mp |\delta \vz^\pm|^2 \rangle .
    \label{eq:Ypm}
\end{equation}
The quantities
$\vct{Y}^\pm$, the 
\emph{Yaglom fluxes}, 
are the only \emph{third}-order structure functions present.
The MHD \vKH equation, Eq.~\eqref{eq:fullform3d}, can then be rewritten as:
\begin{equation}
T^{\pm} + \nabla_\ell \cdot \vY^{\pm}
-2 \nu \nabla_\ell^2 G^{\pm} = -4 \epsilon^{\pm}.
\label{eq:fullform-sym}
\end{equation}
{We will also make use of the sum of these equations which 
represents scale-by-scale conservation
of the total (flow plus magnetic) 
energy rather than that of the Els\"asser energies.  
Writing $T = T^+ + T^-$, and similarly for $Y$ and $G$, we have}
\begin{equation}
     T
   + \nabla_\ell \cdot \vY
   - 2 \nu \nabla_\ell^2  G 
  = 
    -8 \epsDiss.
\label{eq:vKH-epsDiss}
\end{equation}
Below we will refer to the 
$\nabla_{\ell} \cdot \vY^\pm $ terms as the Yaglom term, recalling that 
it represents nonlinear transfer of energy across scales. In
this scale-by-scale energy balance equation, $T$
represents the time rate of change of energy at scales 
smaller than $\ell$, while 
the term involving $G$ is the dissipation at scales larger than $\ell$.

Much of the remainder of this paper will examine various approximations and idealizations in which important information, especially the dissipation rates $\epsilon^{\pm}$ 
 { and $\epsDiss$, } 
can
be extracted easily, and to varying extents, accurately from Eq.~\eqref{eq:fullform-sym}  and \eqref{eq:vKH-epsDiss}.
We emphasize the following properties of this central equation:
(i)
It is exact, subject to the assumptions of spatial homogeneity and incompressibility;
(ii)
It depends on the 3D structure functions of the various terms in vector lag ($\vct{\ell}$) space;
(iii) 
It does not require very large Reynolds numbers;
and
(iv)
It is much more general than the various 
forms of third-order laws \citep{PolitanoPouquet98-vKH,PolitanoPouquet98-grl} that are commonly implemented to estimate $\epsDiss$.

The governing Eq.~\eqref{eq:fullform-sym} is rather versatile as a starting point for determining the bookkeeping of energy at all scales, including its supply from large scales, its transfer across scales 
and its dissipation into heat. When appropriate 
the transfer across scales will be considered to be a \emph{cascade} as will
be discussed more precisely below. 
Equation ~\eqref{eq:fullform-sym}
holds at \emph{every} point in 3D lag space, so if the terms involving 
$T^\pm$, $\vY^\pm$, and $G^\pm$ 
are known at any point, then $\epsDiss$ can be determined.
However, this requires accurate determination 
of first and second derivatives in 
multiple independent lag directions, in general.
Such information is formally available in high resolution 3D simulations, but since such derivatives will be evaluated approximately, averaging results over different $\vct{\ell}$'s---usually at constant $\ell= |\vct{\ell}|$---is useful to achieve accuracy. 
To set the stage for subsequent results we begin with an
illustrative example of this type, 
based on one of our simulations (discussed 
fully in the following section). 
We note that related 
recent studies have also 
implemented a direct evaluation of terms in the MHD \vKH equations, including 
direction averaging \citep{HellingerEA18,AdhikariEA21-reconturb,yang2022pressure}.

 \begin{figure}[ht]
    \centering
    \includegraphics[width = 0.49\textwidth]{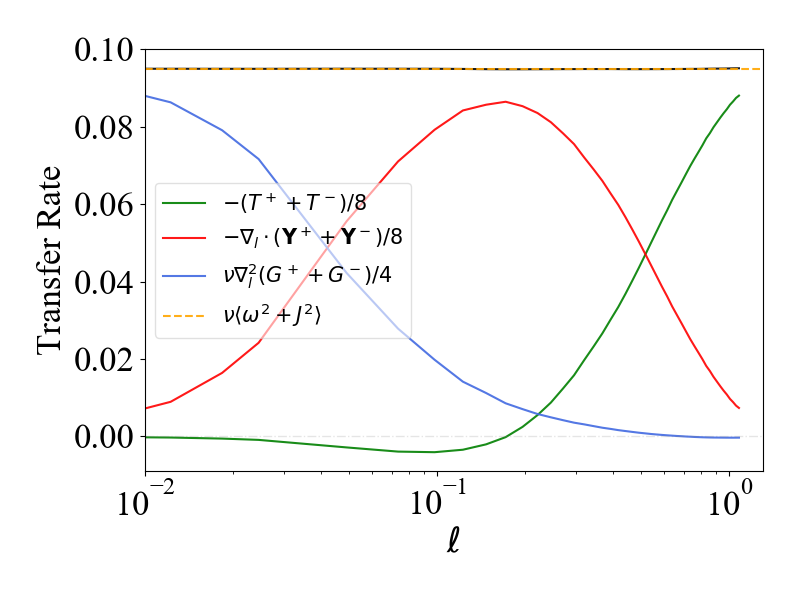}
    \caption{Terms of the \vKH equation, Eq.~\eqref{eq:fullform-sym}, evaluated using the anisotropic 
    ($\vct{B}_0 = 2 \uv{z}$)
    simulation III at $t=3.2$. 
    Here $\frac{\partial}{\partial t} \langle (\delta \vz^{\pm})^2 \rangle = T^{\pm}$, 
    $ \langle \delta \vz^{\mp} |\delta \vz^{\pm}|^2 \rangle = \vY^{\pm}$, 
    and
    $ \langle (\delta \vz^{\pm})^2 \rangle = G^{\pm}$. 
    The green, red, and blue curves correspond to the three LHS terms in Eq.~\eqref{eq:vKH-epsDiss}, averaged over 
    many directions of vector lag  $ \vct{\ell} $.
    Lag is expressed in units of
    the simulation box length ($ 2 \pi $).
    The orange horizontal dashed line indicates the exact dissipation rate $\epsDiss = \nu \langle \omega^2 + J^2 \rangle $. 
    The black curve is the sum of the three (LHS) terms: the large-scale energy supply, the nonlinear transfer rate to smaller scales, and the dissipation at scales larger than $\ell$. As expected, they add up to the total dissipation rate.
    }
    \label{fig:sim3-adir}
\end{figure}
Shown in Figure~\ref{fig:sim3-adir} are the 
{(direction-averaged)}
terms of the \vKH equation, 
Eq.~\eqref{eq:fullform-sym},
plotted versus magnitude of lag for a selected incompressible MHD simulation.  
Upon averaging over directions, the terms in Figure~\ref{fig:sim3-adir} only depend on lag length $\ell$.
It is readily seen that the 
{(appropriate combinations of the)}
$T^\pm$, $\vY^\pm$, and $G^\pm$ terms quite precisely sum to the value of $\epsDiss$. This represents exact conservation of energy; 
specifically, for any chosen length scale,
the sum of the rates of change of energy due to all processes is zero.

It is also apparent that the different terms 
  $T^\pm$, $\vY^\pm$ and $G^\pm$ 
in Eq.~\eqref{eq:fullform-sym} make their principle contribution in different ranges of scales.
In this particular simulation, 
such \emph{separation of scales}
is not perfect; this is discussed further below.
This points to the most commonly encountered 
simplification of scale-dependent energy balance, 
namely the possibility of
an \emph{inertial range}.
In the original hydrodynamic context, \citet{Kol41a}
postulated that an inertial range 
is obtained asymptotically at an infinite Reynolds number.

The onset of an inertial range is expected when the large energy-containing eddies become well-separated from the scales at which dissipation occurs. 
This has been shown in some detail in
high Reynolds number hydrodynamics experiments \citep[e.g.,][]{AntoniaBurattini06}. 
For a true inertial range, 
the Yaglom term $-\nabla_{\ell} \cdot \vY^\pm/4$
is the only {significant} contribution 
to Eq.~\eqref{eq:fullform-sym}
over this range of lag $\ell$ (i.e., this
term is $\gg$ the other terms) and 
we say that the Yaglom term determines the 
\emph{cascade rate}.
Ideally the contribution of 
the Yaglom term is flat over a span of lags.
Fig.~\ref{fig:sim3-adir} exhibits only a hint of the
emergence of such a clear scale separation
and therefore the {phrase} \emph{inertial range}  is only loosely applicable for that simulation. 
Note, however, that the presence or absence of an inertial range does not influence the accuracy of Eq.~\eqref{eq:fullform-sym} in any way.

\subsubsection{Third-order (aka Yaglom) laws}
   \label{sec:third-order}
   
In a well established (i.e., large bandwidth)
inertial range,
both the time variation term ($-T^{\pm}/4$) and the 
scale-dependent dissipation term ($\nu \nabla_\ell^2 G^{\pm}/2$) are small. If these terms become 
vanishingly small over an intermediate scale range 
then---without assuming isotropy---one obtains the simplified equation:  
\begin{equation}
\nabla_\ell \cdot {\bf{Y}}^{\pm} = \grad_{\ell} \cdot \langle \delta \vz^{\mp} |\delta \vz^{\pm}|^2 \rangle = -4\epsilon^{\pm}.  
\label{eq:3d_ez}
\end{equation}
{Such two-term specializations of the \vKH equation are called \emph{third-order} or \emph{Yaglom} laws. 
Herein we refer to Eq.~\eqref{eq:3d_ez} as the \emph{divergence form}, 
or 3D form, of the MHD third-order law.}
A point of emphasis is that to obtain these 
forms requires that the
Reynolds number are large enough that an effectively dissipation-free (inertial) range exists, and that 
the energy content of this range is steady. 
These assumptions are in addition to the requirements of 
homogeneity and incompressibility that are 
inherited from the developments leading to 
Eq.~\eqref{eq:fullform-sym}.
Isotropy is not required.

{\emph{Isotropy.}}
An important historical development 
{was the imposition of isotropy on Eq.~\eqref{eq:3d_ez}, which leads to its}
further simplification.
This was originally done for 
hydrodynamics \citep{Kol41c} and later for MHD            \citep{PolitanoPouquet98-vKH,PolitanoPouquet98-grl}. 
Assuming then that the MHD turbulence is isotropic, 
Eq.~\eqref{eq:3d_ez} may be directly integrated to give a 1D form, or the isotropic form, for the third-order law
             \citep{PolitanoPouquet98-grl,OsmanEA11-3rd}:
\begin{equation}
Y^\pm_\ell = \langle (\uv{\ell}\cdot \delta \vz^{\mp}) |\delta \vz^{\pm}|^2 \rangle  = 
   - \frac{4}{3} \epsilon^{\pm} {\ell} , 
\label{eq:ez}
\end{equation}
sometimes called a ``4/3'' law.
There are also equivalent forms which use just the longitudinal increments---sometimes called the ``4/5'' laws \citep{Kol41c,Frisch,PolitanoPouquet98-vKH}.

{\emph{Direction-averaging.}}
Although isotropy can 
be established a priori
only rarely, 
the isotropic or 1D form of the third-order law is 
{nonetheless still} often used in observational or experimental situations \citep{Sorriso-ValvoEA07,MacBrideEA08}. 
The utility of this approach may be understood better when the technique is supplemented by 
\emph{direction averaging}, carried out in an appropriate way.
This very important and intuitively appealing idea 
has been developed in the hydrodynamics literature 
\citep{NieTanveer99,TaylorEA03}. 
The analogous result for incompressible MHD 
follows by direct extension of the hydrodynamic case. 
Below we present an abbreviated version 
of this straightforward derivation for 
MHD.

We proceed by 
direction-averaging the 
fundamental 
{energy balance}
relation, Eq.~\eqref{eq:fullform-sym}.
Carrying out a full integration over 
all directions,\footnote{Using spherical polar coordinates.} and 
using an overbar to designate  
averaging over the full $4\pi$
solid angle, e.g., 
$\overline{\nabla_{\ell} \cdot \vY^{\pm} } 
= (4\pi)^{-1} \int_S \grad_{{\ell}} \cdot \vY^{\pm}\d\Omega =(4\pi)^{-1} \int_0^{\pi}\int_0^{2\pi} {\nabla_{\ell} \cdot \vY^{\pm}} \,\sin{\theta}
\d\phi \d\theta$,
we find, without loss of generality, that
\begin{equation}
\overline{T^{\pm}} + \overline{\nabla_\ell \cdot \vY^{\pm}}
-2 \nu \overline{\nabla_\ell^2 G^{\pm}} = -4 \epsilon^{\pm}.
\label{eq:directionave}
\end{equation}
It is readily shown that the
angular parts of the $\nabla_\ell $ operators do not 
contribute when the averaging 
is taken into account.
Some details are provided in Appendix~\ref{appA}.
The direction-averaged \vKH equation 
becomes:
\begin{eqnarray}
\overline{T^{\pm}}
+ \frac{1}{\ell^2}\frac{\d}{\d\ell} [\ell^2 \overline{Y^{\pm}_\ell}]
- \frac{2 \nu}{\ell^2} \frac{\d }{\d\ell}
\left[\ell^2 \frac{\d}{\d\ell} \overline{G^{\pm}}
\right] &=& -4 \epsilon^{\pm} \nonumber\\
{\rm or, }\hspace{20pt}  
\overline{T^\pm} + {\cal D}^{(1)}_\ell\overline{Y^{\pm}_\ell}
- 2\nu {\cal D}^{(2)}_\ell \overline{G^{\pm}} 
&=& -4 \epsilon^{\pm}
\label{eq:avereduced}
\end{eqnarray}
where each term on the left hand side depends on the scalar lag $\ell = | \vct{\ell}|$.
In Eq.~\eqref{eq:avereduced}, for convenience of presentation later, 
we have introduced the shorthand notation
$ {\cal D}^{(1)}_\ell \equiv \frac{1}{\ell^2}\frac{\d}{\d\ell} [\ell^2 *] $ for the radial contribution to the divergence operator, 
and 
${\cal D}^{(2)}_\ell \equiv  \frac{1}{\ell^2} \frac{\d }{\d\ell}
\left [\ell^2 \frac{\d}{\d\ell} *\right ]$
for the radial part of the Laplacian operator, with dummy arguments $*$, and both defined in the lag space. 
We emphasize this explicitly 
to distinguish the form of Eq.~\eqref{eq:fullform-sym}
which involves three dimensional (3D) vector operations,  
while
Eq.~\eqref{eq:avereduced}
involves only the reduced dimensional 1D operations. 

This direction-averaged form of the \vKH equation, Eq.~\eqref{eq:avereduced}, remains quite general requiring only spatial homogeneity and incompressibility.  
If the system is also time stationary, or if suitable time averaging is performed \citep{TaylorEA03}, the first term, $\overline{T^{\pm}}$, may be safely neglected.  
Following the usual arguments, when the 
Reynolds number is sufficiently large ($\nu$ sufficiently small),
the dissipative term involving $\overline{G^{\pm}}$
is also negligible over an intermediate range, which becomes the inertial range. In that case the remaining 
ordinary differential equation is
\begin{equation}
{\cal D}^{(1)}_\ell\overline{Y^{\pm}_\ell}
=\frac{1}{\ell^2}\frac{\d}{\d\ell} [\ell^2 \overline{Y^{\pm}_\ell}]=-4 \epsilon^{\pm},
\end{equation}
which immediately integrates 
to the 
\emph{direction-averaged  third-order law}
(cf.\ Appendix Eq.~\eqref{thirdorderaveraged}) 
\begin{equation}
    \overline {Y_{\ell}^{\pm}} = -\frac{4}{3} \epsilon^{\pm} \ell.
\label{eq:1Dform}
\end{equation}
{Although this is structurally identical to the 1D form third-order law associated with an inertial range in \emph{isotropic} cases, Eq.~\eqref{eq:ez},
we emphasize that 
 there are several important distinctions between Eqs.~\eqref{eq:ez} and \eqref{eq:1Dform}.  The former holds at any (inertial range) vector lag $\vct{\ell}$ but only for isotropic turbulence. 
 The latter holds for \emph{any} rotational symmetry (or lack thereof) but requires averaging  (the full $4 \pi$) 
 solid angle.}

The third-order  (Yaglom) laws for hydrodynamics and for MHD 
{should be} 
applied in situations in which one may reasonably assume that the conditions 
leading to these relations are actually attained. However, in general the 
most frequently quoted conditions ---
time-stationarity, homogeneity in space,
and high Reynolds numbers --- may not always 
hold and a pristine inertial range may not 
appear. In such cases the range over which a third-order law might be applied 
may be ``polluted'' by other terms in Eq.~\eqref{eq:fullform-sym}.

The exact statement of energy conservation in Eq.~\eqref{eq:fullform3d}
or \eqref{eq:fullform-sym} provides a complete specification of the energy 
balance in homogeneous turbulence when evaluated over 
arbitrary regions of the (vector) lag space.
When the full energy balance cannot be computed, 
it is common practice to resort (sometimes without demonstrating justification) to more compact 
third-order laws --- such as 
    Eqs.~\eqref{eq:3d_ez}, \eqref{eq:ez}, and \eqref{eq:1Dform} ---
all of which require that the time dependent terms
$T^\pm$ and dissipative terms $G^\pm$ 
are negligible 
for a useful region of lag space. 

Below we provide several examples of different 
approaches to approximately measure the inertial range transfer 
rate (sometimes loosely called the 
\emph{cascade rate})
using MHD simulation data:
~\\

\noindent \emph{Method I.  }
The unidirectonal 
1D form,
Eq.~\eqref{eq:ez}, is evaluated for a fixed lag direction, over a range of lag magnitudes. This is  suitable for isotropic systems. 
The choice of direction is clearly not unique \textbf{but should not matter for a truly isotropic system.} There are several variations:
Eq.~\eqref{eq:ez}  can be evaluated over a range of lags with a linear fit (through the origin) giving $\epsilon^\pm$. Alternatively the equation can first be divided by $\ell$ and the 
result plotted, allowing esitmation of $\epsilon^\pm$ from a suitable flat range. 
~\\

\noindent \emph{Method II.}
The 3D (or, divergence) form, Eq.~\eqref{eq:3d_ez},
can be used to compute $\epsilon^\pm$ when 3D lag-space derivatives can be reliably determined in several directions. 
This method is based on
a direct evaluation of (derivatives of) the $\vY^\pm$ terms in the 
\vKH equation \eqref{eq:fullform-sym}
and is fully general in terms of turbulence symmetries when 
inertial range conditions are obtained. 
No direction averaging is required although averaging
may reduce statistical inaccuracies. 
In this paper, the only occasion that the 3D lag-space \emph{derivatives} are calculated is when obtaining the curves shown in 
    Figure~\ref{fig:sim3-adir}.  
~\\

\noindent \emph{Method III}. 
The direction-averaged 1D form,   
Eq.~\eqref{eq:1Dform} is exact when 
integrated over the full spherical domain of lags, 
provided that inertial range conditions are established.
However a full integration over direction
may not be feasible with available measurements.
In practice a discretized approximation to the continuum average is likely to be needed.  This may be obtained, for example, 
by calculating $ Y^\pm_\ell $ for each of the (limited number of) available directions of $ \vct{\ell} $ and then forming the appropriately weighted average of these, see for example Eq.~\eqref{discret}. ~\\

All three of the above Methods, being based on third-order laws, require the existence of an inertial range, at least approximately.
Otherwise, and more generally, when the 
 $T^\pm$ and $G^\pm$ terms are significant, it is appropriate to use more complete forms of the \vKH equation, such as Eq.~\eqref{eq:fullform-sym} or ~\eqref{eq:avereduced}.

\subsection{Observational Approaches and  Limitations}\label{subsec:observe_limits}

Typical solar wind studies of turbulence are carried out with single spacecraft 
measurements and in a high speed flow for which time correlations
can be interpreted as spatial correlations with reasonable accuracy \citep{Jokipii71}. 
In these circumstances observational analyses
typically 
compute a cascade 
rate  
by employing 
1D forms of the third-order law or its generalizations.
In space plasma measurements, it is difficult to obtain the time variation and the dissipative terms in the \vKH equation Eq.~\eqref{eq:fullform3d} or \eqref{eq:fullform-sym}, 
and therefore
only the \divYpm
terms (or really their integrals) can be calculated
using \emph{in situ} data from single spacecraft. 

Most frequently Method I,
    Eq.~\eqref{eq:ez},
is employed for cascade rate estimation in solar wind  
\citep{MacBrideEA-sw11,Sorriso-ValvoEA07,MacBrideEA08}.
Implicit in this approach is the 
assumption of isotropy, although 
the accuracy of this approximation has rarely been 
demonstrated. 
There have been attempts to 
adapt the 1D forms 
in order to refine the method 
\citep{StawarzEA09,CoburnEA15} by making various 
assumptions about the structure and symmetry of the Yaglom flux. 

The 1D forms, essentially Eq.~\eqref{eq:ez}, 
have also been applied in the 
earth's magnetosheath \citep{BandyopadhyayEA21-mtail}
and in Parker Solar Probe data near perihelia \citep{BandyopadhyayEA20-PSP3rd} where turbulence is much more intense than at 1\,au. 
It is noteworthy that there is considerable recognition that averaging is required, 
although this typically takes the form of a requirement for large data volumes and 
large data sets
\citep{Dudok04,PodestaEA09-3rd} rather than a requirement for  averaging over lag directions.

In single spacecraft measurements (with Taylor hypothesis)
it is feasible to improve accuracy 
by averaging several measurements 
\citep{MacBrideEA08}.
However, this averaging method is still not considered as the 3D form Eq.~\eqref{eq:3d_ez}, and generally the weighting of different directions has not been considered. 
In particular,
keeping in mind the results summarized in
the previous section,
finite sampling strategies 
generally do not guarantee a 
uniform distribution of lag directions on a sphere. 
In this regard 
a very important development is the recognition that the 
Yaglom flux varies systematically over the direction
relative to the mean magnetic field
\citep{VerdiniEA15}.
However the assembly of measurements 
into a proper averaging over the unit sphere
-- essentially the result of \cite{NieTanveer99} in hydrodynamics extended to the 
MHD third-order law \citep{PolitanoPouquet98-vKH} as in Eq.~\eqref{eq:1Dform}
-- has apparently not been fully
appreciated in previous MHD and space physics studies. 
There has been at least one study \citep{OsmanEA11-3rd}
employing multispacecraft Cluster data
that accumulated the normal flux over a sphere 
in lag space, approximately carrying out the operations implied by  Method III, Eq.~\eqref{eq:1Dform}.
Such datasets are infrequently available in the solar wind.

As a consequence, it is crucial to know how accurately 
one might estimate the cascade or dissipation rate 
when computed from Method I, the
1D form Eq.~\eqref{eq:ez}, 
in various situations.
This may be challenging 
in the solar wind, where the existence of a strong global magnetic field implies significant anisotropy.
Proper direction averaging may not always be possible, unless 
very large ensembles are considered, as in, for example, recent ensemble average computations of 
correlation function that employed
years of data and proper normalization of individual samples
\citep{RoyEA21}. However the intent of cascade
rate estimation is often to understand more \emph{local}  
conditions, so the emphasis may be on 
very much more local averaging. 
Such cases are 
severely constrained by the availability of 
single spacecraft data and the number of directions 
relative to the mean magnetic field that can be sampled.
For an examination of the distribution of 
flow-magnetic field directions 
at 1\,au, see the analysis of this question
based on the MMS Turbulence Campaign in the solar wind 
by \citet{ChasapisEA20}.
The remainder of this paper 
is largely devoted to exploring accuracy of energy transfer (cascade) rate
measurements using \vKH equations 
and third-order laws in various forms.

\section{Simulations}\label{sec:sims}

In order to study the energy cascade rate in MHD turbulence using third-order structure functions, we analyze data from several incompressible 3D MHD turbulence simulations. Key parameters of the simulations employed are shown in Table~\ref{tab:sim_run}. 
For all simulations in the table, the domain is
a three-dimensional periodic box with sides of length $2 \pi$, and the
MHD equations are solved 
using a Galerkin spectral method \citep{OrszagPatterson72,OughtonMatt20}.
Each simulation is initialized with the condition that the fluctuation magnetic energy and fluid flow energy are equal, such that $E_m = E_f = 0.5$. Also, the viscous and resistive dissipation coefficients, $\nu$ and $\mu$, are set equal. These parameters act, respectively, as
reciprocal Reynolds number and magnetic Reynolds number.
The normalized cross helicity, 
  $ \sigma_c =  \langle \vct{v}\cdot \vct{b} \rangle /  (E_f + E_m) $,
is known to be a significant factor in evolution of turbulent MHD \citep{PouquetEA86}.
Run III has a
substantial value of $\sigma_c$,
and this case will be qualitatively 
contrasted with the lower cross helicity case in run II.
A full scan of parameters such as $\sigma_c$ will not be attempted in this 
study.
Runs II and III are undriven and anistropic, with distinct values of $B_0$.
Run I is isotropic and driven; this case is included to minimize time dependent effects 
while examining residual transient dependence 
on direction, an effect that will be discussed further below.

\begin{figure}[htb]
    \centering
    \includegraphics[width = 0.3\textwidth]{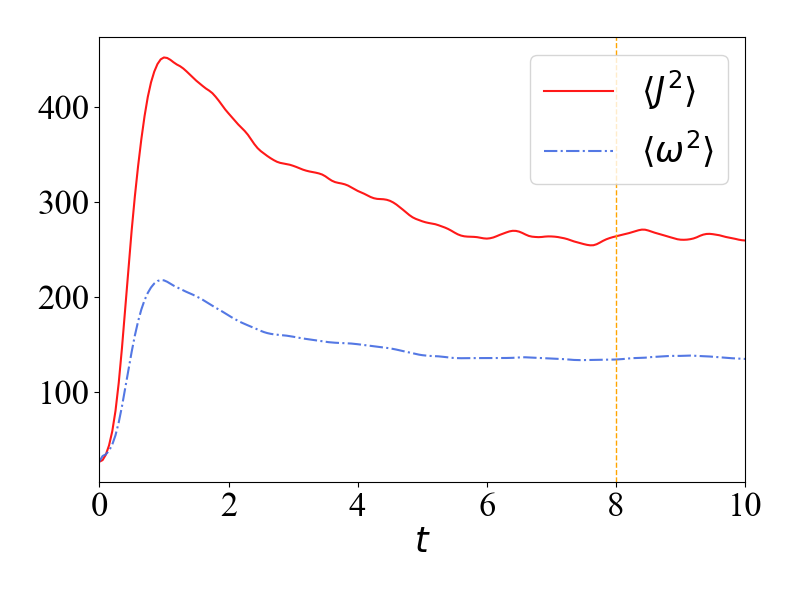}
    \includegraphics[width = 0.3\textwidth]{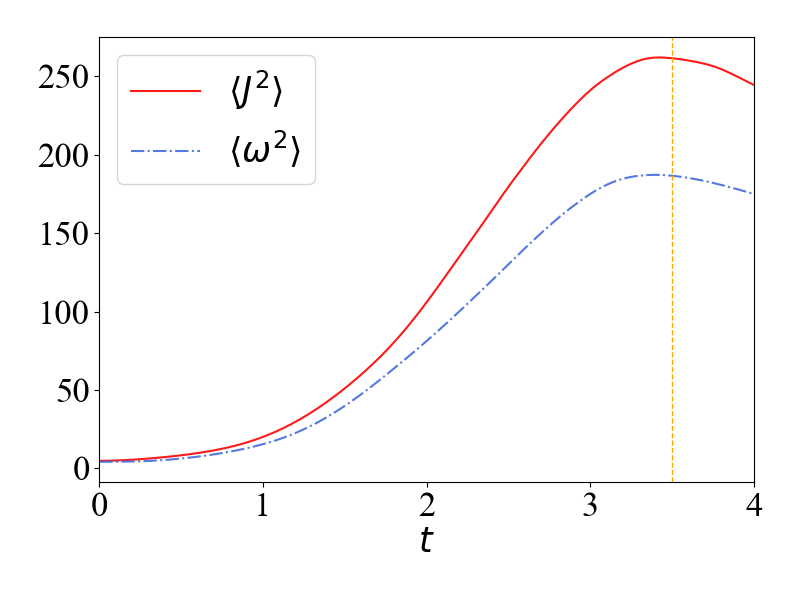}
    \includegraphics[width = 0.3\textwidth]{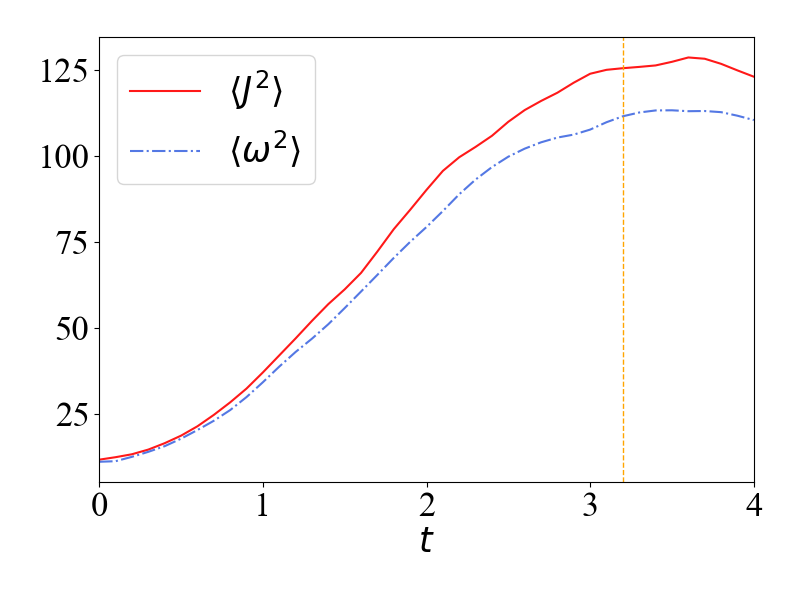}
    \caption{Time evolution of $\langle J^2 \rangle$ and $\langle \omega^2 \rangle$ for runs I (left), II (middle), and III (right). The vertical orange line indicates the time for performing analysis. }
    \label{fig:Et}
\end{figure}

The analyses presented here 
are carried out at the simulation times indicated in Fig.~\ref{fig:Et}, which shows time evolution of mean square current density and mean square vorticity for each run.  
We can compute 
the exact value of the dissipation rate, $ \epsDiss $, using Eq.~\eqref{eq:eps-diss}.
Table~\ref{tab:sim_run} lists the values of $\epsDiss$ for each simulation at the time our analysis is performed.

\begin{table}[ht]
\centering
\begin{tabular}{||c c c c c c c c c c c||} 
 \hline
 Run & Type & symmetry & Resolution (3D) & $B_0$ & $\delta b/B_0$ & $k$ range & $\nu=\mu $ & $\sigma_c$ & $\epsDiss$ & $\ell_\text{diss}=(\nu^3/\epsDiss)^{1/4}$ \\ [0.5ex] 
 \hline\hline
 I & driven & isotropic & 512 & 0 & - & 3-5 & $2.0\times 10^{-3}$ & -0.05 & 0.795 & 0.010\\ 
  II & decaying & anisotropic & 1024 & 1 & 1  & 1-3 & $4.0\times 10^{-4}$ & 0.09 & 0.179 & 0.0043\\  
 III & decaying & anisotropic & 1024 & 2 & 0.5 & 1-5 & $4.0\times 10^{-4}$ & 0.7 & 0.0948 & 0.0051\\ [1ex]
 \hline
\end{tabular}
\caption{
Simulation parameters. 
The global magnetic field $\vct{B}_0$ is in  the $\uv{z}$ direction. The rms magnetic fluctuation is $\delta b  = \sqrt{\langle \vct{b}^2 \rangle}$, and
$\delta b/B_0$ measures the initial
relative strength of the fluctuations. 
Column ``$k$ range'' indicates the wavenumber forcing band (Run I) or the initial conditions band (Runs II and III). 
Viscosity $\nu$ equals resistivity $\mu$ for each simulation. The normalized cross helicity $\sigma_c $, 
energy dissipation rate $\epsDiss $,
and dissipation scale $\ell_\text{diss} $
are computed at the respective times of analysis indicated  in Fig.~\ref{fig:Et}. 
 }
 \label{tab:sim_run}
\end{table}

In following sections, we carry out several types of analyses centered around the strategy of employing the third-order structure function to estimate the dissipation rate in the simulation.
The exact dissipation rate, $\epsDiss$ from Eq.~\eqref{eq:eps-diss}, is used to study the accuracy of these various strategies.

\section{Results: 1D form third-order law} \label{sec:results_1Dlaw}

In this section  we apply 
  Eq.~\eqref{eq:ez},
a simplified and widely used 
1D form of the third-order law 
\textbf{referred to as \emph{Method I},
to simulation data obtained from} 
from both isotropic (\S\ref{subsec:isotropic})
and anisotropic cases (\S\ref{subsec:anisotropic}). 
We examine the inertial ranges and dissipation rates estimated from this 1D form. We will consider estimates based on individual directions as well partial averages over directions, but we do not 
here attempt a full integration over solid angle, which is a requirement for \emph{Method III}. 

\subsection{Isotropic case}  \label{subsec:isotropic}



\begin{figure}[ht]
    \centering
    \includegraphics[width = 0.4\textwidth]{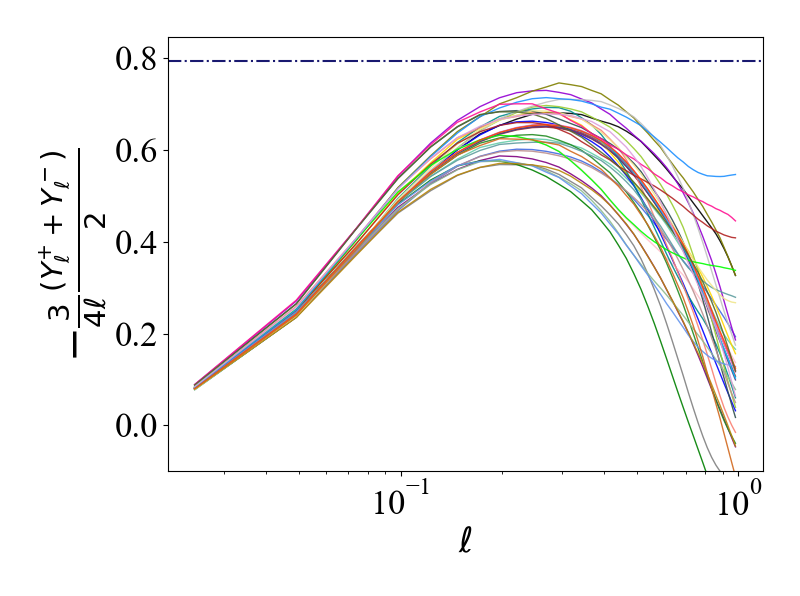}
    \caption{Estimating the dissipation rate using Method I, the 1D form third-order law (Eq.~\eqref{eq:ez}), applied to data from isotropic simulation I.
    Recall $Y^\pm_\ell = \langle (\uv{\ell} \cdot \delta \vz^\mp) |\delta \vz^\pm|^2 \rangle$. 
    Different curves represent results of 36 different lag directions uniformly distributed on a sphere. Each curve represents the average of the $Y_{\ell}^+/\ell$ and $Y_{\ell}^-/\ell$ terms for a fixed lag direction. The dark blue dashed horizontal line (at 0.795) indicates the actual energy dissipation rate.
    A standard procedure is to assume that 
    the peak values provide estimates of the dissipation rate and the corresponding $\ell$ values locate the middle of the inertial range.}
    \label{fig:ssb0-1d-dd}
\end{figure}

We consider the driven $B_0 = 0$ isotropic simulation I.  Figure~\ref{fig:ssb0-1d-dd} displays the curves for $(Y^+_\ell+ Y^-_\ell)/ \ell $ as a function of $\ell$, individually computed
for each of 36 selected directions uniformly distributed on a sphere.
The 3D trilinear interpolation method \citep{bai2010comparison} is used to calculate magnetic and velocity fields not located on grid points. Each curve represents a certain lag direction, and the value along the $y$ axis is the average of $-3Y_{\ell}^+/(4\ell)$ and $-3Y_{\ell}^-/(4\ell)$. 
The peak value of $-3(Y_{\ell}^++Y_{\ell}^-)/(8\ell)$
is used to estimate the dissipation rate. 
We observe that for all the curves the peak values are smaller than the actual dissipation rate, $\epsDiss$,
which is explained in \S\ref{subsec: vkh_results}.
It is also evident in Fig.~\ref{fig:ssb0-1d-dd} that the inertial ranges associated with the different lag directions
are broadly consistent in extent, with some variation in the peak values.
This indicates that at the instant of time of this analysis, even this nominally isotropic simulation admits some degree of variation over directions.
It is reasonable to suppose that directional averaging might improve the
estimates of dissipation rate in this case; 
we will take up this discussion in a later section. 

\subsection{Anisotropic case}\label{subsec:anisotropic}

\begin{figure}[ht]
    \centering
    \includegraphics[width = 0.4\textwidth]{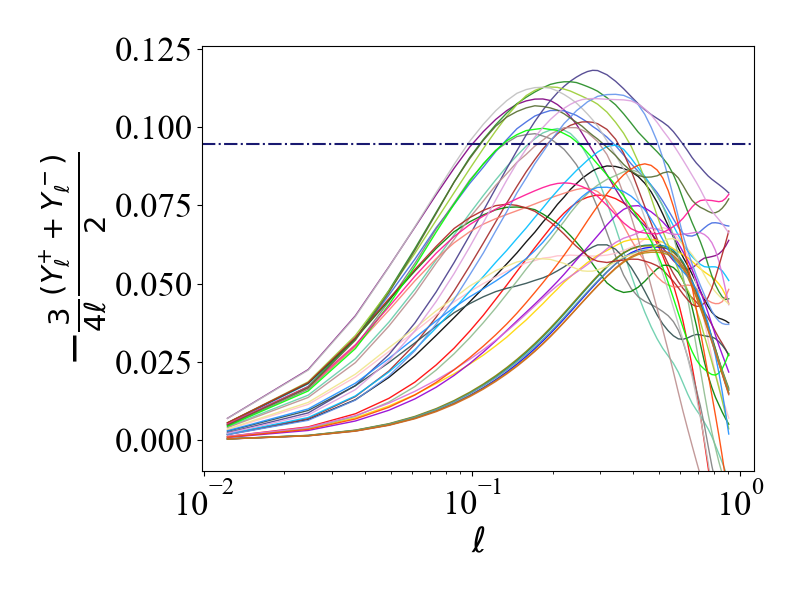}
    \includegraphics[width = 0.4\textwidth]{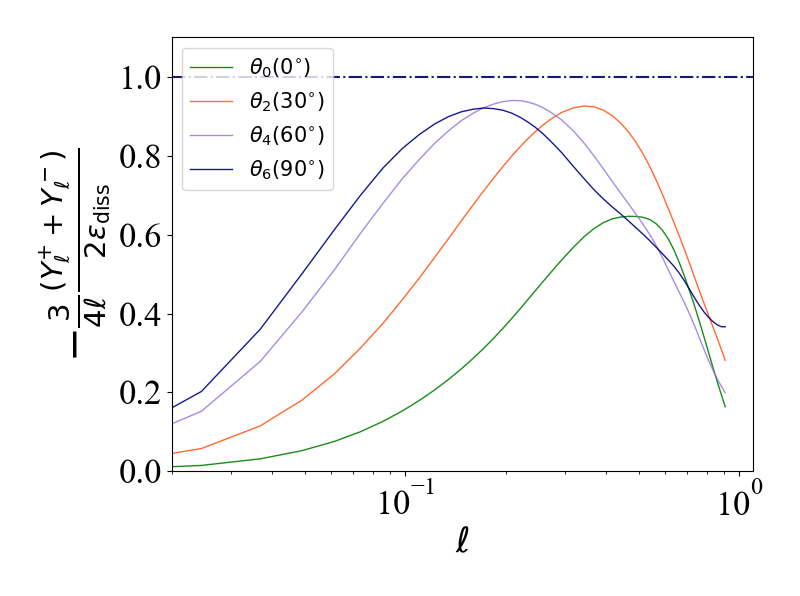}
    \caption{A collection of estimates of dissipation rate 
    using Method I, the 1D form third-order law
    (Eq.~\eqref{eq:ez}), 
    applied to data from the high $\sigma_c$, $\vct{B}_0 = 2\uv{z}$ 
    anisotropic simulation III. 
    {Left: Each curve represents the average of the $Y_{\ell}^+/\ell$ and $Y_{\ell}^-/\ell$ terms for a fixed lag direction. Dashed horizontal line 
    indicates the actual energy dissipation rate.
    Peak values estimate the dissipation rate and (horizontal) location of a peak roughly locates the middle of the inertial range.
    Right: 
        same data as left panel, normalized by $\epsDiss$, and averaged
        over azimuthal angle $\phi$.}
    Each curve is for 
    a fixed polar angle $\theta$, and averaged over six 
    equally spaced azimuthal angles with $\Delta \phi = \frac{\pi}{3}$. 
    }
    \label{fig:lsb2-1d-dd}
\end{figure}

The anisotropic simulations we consider have a mean magnetic field, $B_0 \uv{z} $, with $B_0 = 1$ or 2
and differing cross helicities
(see Table~\ref{tab:sim_run}).
In this section, we only consider the anisotropic simulation III.
To examine 
how the lag direction impacts the Yaglom term in anisotropic MHD, we 
evaluate the dissipation rate in 
simulation III
using Method I, i.e., the 1D form of the third-order law, Eq.~\eqref{eq:ez}.
Separate estimates are made 
using lags in each of 36 directions, uniformly spaced in 
co-latitude and azimuthal angles ($\Delta \theta = \frac{\pi}{6}$ and $\Delta \phi =\frac{\pi}{3}$).  
The left panel of Fig.~\ref{fig:lsb2-1d-dd} 
demonstrates that the peak values associated 
with these different lag directions 
occur at different lags.
The levels of the maxima also vary, in some cases 
exceeding the true dissipation rate.
Similarly, the `inertial ranges' associated with the different lag directions also vary in bandwidth and position.

In order to further study the effect of 
partial averaging in the 
presence of a global magnetic field, 
we group the lag directions by their corresponding polar angle $\theta$. 
Recall that the global magnetic field is in $+\uv{z}$ direction. For each polar angle, we 
  {average the $Y^\pm_\ell / \ell $}
terms over 6 azimuthal directions (Fig.~\ref{fig:lsb2-1d-dd}, right panel). We observe that 
the peak of each curve shifts to smaller lags as  $\theta$ increases. 
Moreover, 
 \textbf{the peak value associated with} 
the quasi-parallel 
case ($\vct{\ell} \parallel \vct{B_0}$ and $\theta=0$) 
provides a significantly lower estimate of $\epsDiss$ 
than do the larger $\theta$ cases, which have peak values comparable to the actual dissipation rate. This reflects the well-known 
fact that energy transfer proceeds more rapidly perpendicular
to an applied magnetic field \citep{ShebalinEA83}.   

\begin{figure}[ht]
    \centering
    \includegraphics[width = 0.4\textwidth]{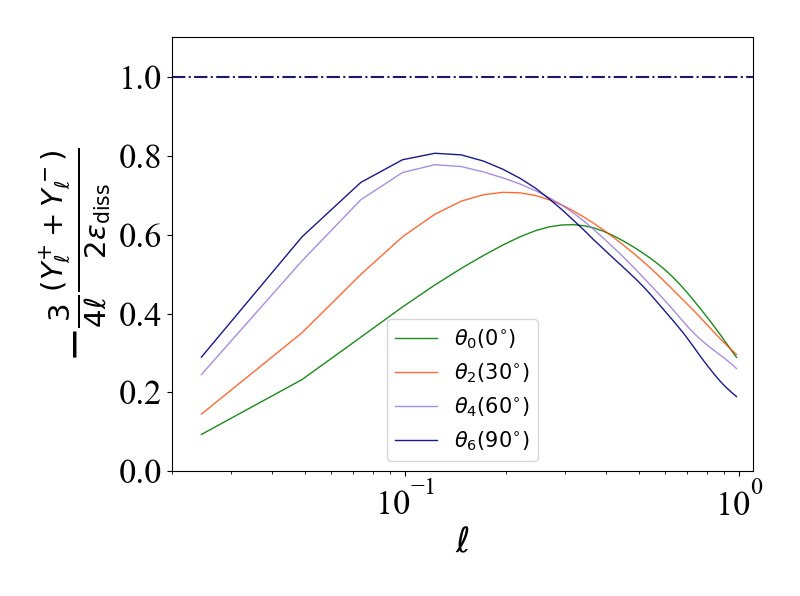}
    \caption{Evaluating the dissipation rate using the 1D form third-order law (Eq.~\eqref{eq:ez}, Method I) for the low $\sigma_c$, $B_0 = 1$ anisotropic simulation II. 
    Curves correspond to lag directions with the indicated polar angle $\theta$ and averaged over six 
    equally spaced azimuthal angles, $\phi $. Each curve represents the average of the corresponding $ Y_{\ell}^\pm / \ell $ terms,
    normalized by the true dissipation rate. 
    }
    \label{fig:nsb1-1d}
\end{figure}

An analysis similar to that of 
Fig.~\ref{fig:lsb2-1d-dd} 
is 
shown in Fig.~\ref{fig:nsb1-1d} 
for simulation II, which has low $\sigma_c$ and 
$B_0 = 1$.
Here, once again, 
larger $\theta$ values, 
more strongly perpendicular lag directions,
are associated with 
inertial range behavior 
found at smaller lags. 
It is interesting to examine the degree of anisotropy of the energy transfer by looking at the 
disparity of the cascade rate estimates at varying angles in the two cases, 
simulation II with weaker $B_0$ and 
smaller $\sigma_c$, and simulation III with stronger $B_0$ and larger $\sigma_c$. 
The ratio of strongest estimate to weakest over angles is actually 
sightly greater in 
simulation II 
(ratio $\sim 1.4$) than in simulation III
(ratio $\sim 1.3$),
even though simulation III has the stronger mean magnetic field. 
Superficially this result appears to be anomalous; however, the 
significant contrast in cross helicity ($\sigma_c$) values
is a likely explanation, as this is another factor that can 
influence nonlinear timescales, cascade strength, and anisotropy.  

\subsection{Comparison between isotropic and anisotropic cases using 1D form}\label{subsec:1D_comps}

{The above Method I results for estimating $\epsDiss$ indicate clear differences between the isotropic and anisotropic situations.}
In each system, we chose 36 different lag directions, distributed uniformly on a sphere, and employed the 1D form of the third-order law (Eq.~\eqref{eq:ez}) to calculate $ (\epsilon^+ + \epsilon^-)/2$ for each direction. 
For 
the driven (statistically steady) isotropic simulation
we found that the inertial range and also the corresponding energy cascade rate 
determined this way are roughly independent of the lag directions (Fig.~\ref{fig:ssb0-1d-dd}), as expected for isotropy.
For the two anisotropic cases different lag directions 
vary much more in 
terms of the cascade rate estimates as well as 
location and bandwidths of the suggested inertial range(s).
The conclusion is that 
it is difficult to accurately determine the energy cascade rate of an anisotropic system using the 1D form third-order law (Eq.~\eqref{eq:ez}), especially with one or a small 
number of computed lag directions. 
For an isotropic system the situation is somewhat better, although some 
modest variation in estimated cascade rate is seen in varying lag directions.

\section{Results: Full Directional Averaging}   \label{sec:results_3Dlaw}

Given the variability we have seen in the above numerical experiments in both isotropic and anisotropic cases, 
we expect to obtain improved results 
starting from either the
divergence (3D) form Eq.~\eqref{eq:3d_ez} (Method II),
valid within a well-defined inertial range, 
or from the \vKH equation Eq.~\eqref{eq:fullform3d} or \eqref{eq:fullform-sym}, 
which is broadly applicable even when
an inertial range is not present. 
Another strategy,
which we now explore, 
is to 
compute the dissipation rates $\epsilon^{\pm}$ using
Method III, the fully direction-averaged 1D form of the third-order law, 
Eq.~\eqref{eq:1Dform}.\footnote{We already showed results in Fig.~\ref{fig:sim3-adir} obtained from 
the  \vKH equation (Eq.~\eqref{eq:fullform3d} or \eqref{eq:fullform-sym}) which was also direction-averaged as in Eq.~\eqref{eq:directionave}. This averaging over directions in Fig.~\ref{fig:sim3-adir} was not required but 
increased accuracy of the computation.} 

As emphasized above, 
based on the generalization of the result of 
\citet{NieTanveer99} to the MHD
\vKH equation, one finds that direction-averaging
can reduce the problem to the 1D integration over the full $4\pi$ solid angle, see for example, Eq.~\eqref{eq:avereduced}. 
We can then consider just the radial component if the chosen lag directions completely cover the sphere (details of equations and discretization method can be found in Appendix~\ref{appA}).
In particular, when an inertial range 
is present, the direction-averaged 
MHD Yaglom law Eq.~\eqref{eq:1Dform} emerges as an exact result.

Here we proceed numerically, 
employing a discretization method 
like Eq.~\eqref{discret} to calculate the direction-averaged 1D form third-order law 
Eq.(\ref{eq:1Dform})
in the inertial range, and a similar method for the 
direction-averaged
\vKH equation Eq.~\eqref{eq:avereduced} (details in Appendix~\ref{appA}). 
Again, 3D trilinear interpolation is used for values of magnetic field and velocity not located on grid points. In order to get a uniform distribution, at any fixed lag length, we vary the direction of the lag. Let $\theta$ be the polar angle, and $\phi$ be the azimuthal angle, in such case, we keep $\Delta \theta$ and $\Delta \phi$ to be fixed, which are $\pi/12$ and $\pi/6$, respectively, for simulations with resolution 1024 ($\pi/8$ and $\pi/4$ for simulations with resolution 512).

\subsection{Method III: Direction-Averaged 1D Form Third-Order Law} \label{subsec:3d_ini}

\begin{figure}[ht]
    \centering
    \includegraphics[width = 0.49\textwidth]{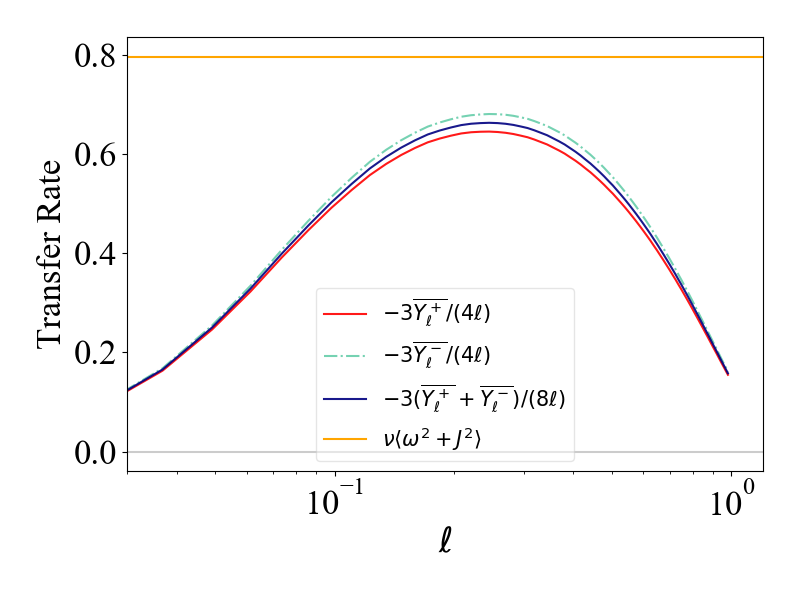}
    \caption{Evaluating dissipation rates for isotropic simulation I using Method III, the direction-averaged 1D form third-order law, Eq.~\eqref{eq:1Dform}. Here the 
    direction-averaged $Y^\pm_\ell = \langle (\uv{\ell}\cdot \delta \vz^{\mp}) |\delta \vz^{\pm}|^2 \rangle$
    term is designated as 
    $\overline{Y^{\pm}_{\ell}}$.
    The dark blue curve is the average of the two $\overline{Y^{\pm}_{\ell}}$ terms, and the orange horizontal line indicates the exact dissipation rate. }
    \label{fig:y-3d-b0ss}
\end{figure}

\begin{figure}[ht]
    \centering
    \includegraphics[width = 0.49\textwidth]{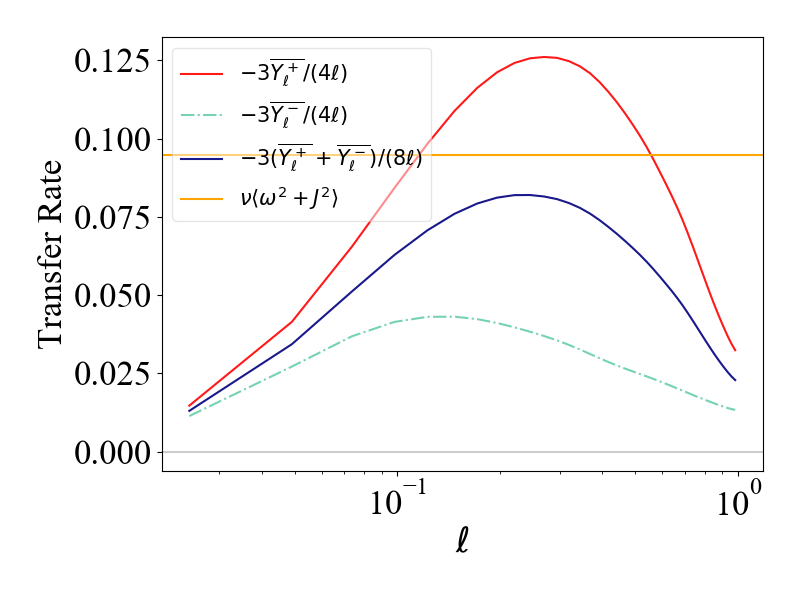}
    \caption{Method III, direction-averaged 1D  form third-order law (Eq.~\eqref{eq:1Dform}),
    applied to data from anisotropic simulation III, which has high $\sigma_c = 0.7$ and $B_0 = 2$. The same color code as Fig.~\ref{fig:y-3d-b0ss} is used.
    }
    \label{fig:yg-3d}
\end{figure}

In Figs.~\ref{fig:y-3d-b0ss} and \ref{fig:yg-3d}, we show the directional average of 1D form third-order law, with its assumption of a well-defined inertial range, using data from simulations I and III. Recall that 
simulation III has a high cross helicity, leading to a large difference between ‘+’ and ‘-’ terms in Fig.~\ref{fig:yg-3d}. The dark blue curve is the average of the minus and plus terms. 
We see that the dissipation rate computed
from the direction-averaged 1D form 
is slightly smaller than the exact dissipation rate in both 
isotropic and anisotropic cases.
This indicates a relatively minor role of the time variation term and the dissipative term in the \vKH equation for both cases.

\subsection{ Direction-Averaged von K\'arm\'an--Howarth Equation} 
\label{subsec: vkh_results}

We now demonstrate estimation of energy transfer rates
using all relevant terms in the direction-averaged \vKH equation Eq.~\eqref{eq:avereduced}.
Note that in a 
driven system, 
the \vKH equation Eq.~\eqref{eq:fullform3d}
should be extended to include
a large-scale forcing term $\langle \delta \vz^\pm \cdot \vct{F}_\ell \rangle$
on the right hand side. The associated direction-averaged form 
can be written as 
\begin{equation}
    \overline{T^\pm} 
  + {\cal D}^{(1)}_\ell\overline{Y^{\pm}_\ell}
  - 2\nu {\cal D}^{(2)}_\ell \overline{G^{\pm}} 
= 
  -4 \epsilon^{\pm} + \overline{\langle \delta \vz^\pm \cdot \vct{F}_\ell \rangle}
\label{eq:averforced}
\end{equation}

\begin{figure}[ht]
    \centering
    \includegraphics[width = 0.49\textwidth]{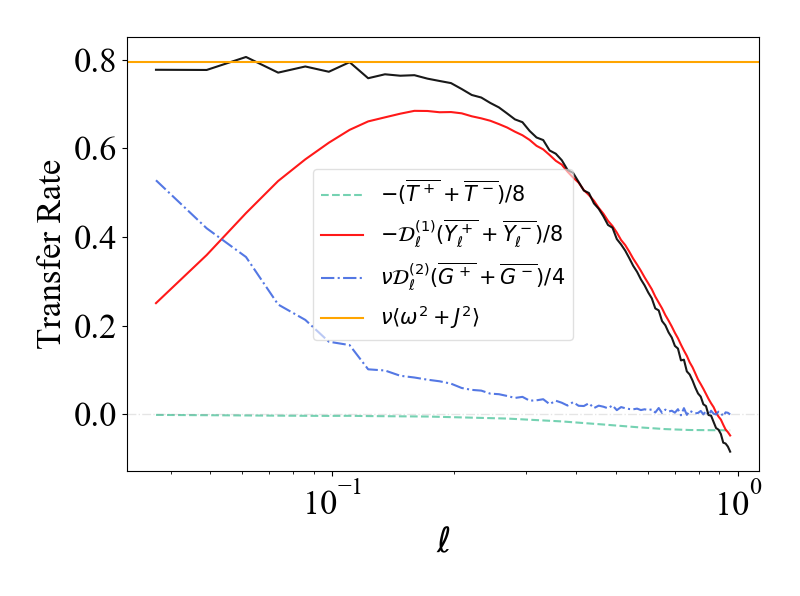}
    \caption{Terms of the direction-averaged
    \vKH equation Eq.~\eqref{eq:avereduced} in the driven isotropic simulation I at $t=8$.
    The orange horizontal line indicates the actual dissipation rate.
    The black curve, representing the total transfer rate, is the sum of the three lines representing the $Y$, $G$ and $T$ terms. The forcing term, which contributes to the transfer rate on large scales, as in 
    Eq.~\eqref{eq:averforced}, is not plotted.}
    \label{fig:ygt-all-I}
\end{figure}

This analysis is first carried out for 
simulation I, a driven isotropic system.
Fig.~\ref{fig:ygt-all-I} displays
these direction-averaged contributions to 
the \vKH equation, 
omitting the forcing term. 
We see that at small scales the sum of the (direction-averaged)
time variation term ($-\overline{T^\pm}$), 
cascade term ($ -{\cal D}^{(1)}_\ell\overline{Y^{\pm}_\ell}$), 
and dissipative term ($2\nu {\cal D}^{(2)}_\ell \overline{G^{\pm}}$) 
add up to the actual dissipation rate; evidently
the driving force term does not play a role at these scales. 
On the other hand, at
large scales, the forcing term (not shown) 
is dominant, and
the sum of the other three terms drops as $\ell$ increases. 
One may notice that when the Yaglom term
${-\cal D}^{(1)}_\ell(\overline{Y^+_\ell}+\overline{Y^-_\ell})/8$
is at its peak of \textbf{$\sim 86\%$ of $\epsDiss $,}
the dissipative term $\nu {\cal D}^{(2)}_\ell (\overline{G^+}+\overline{G^-})/4$ is $ \sim 12\%$ of $\epsDiss$, 
which is small although
not quite negligible.
Comparing this to the results shown in
  Fig.~\ref{fig:y-3d-b0ss}, obtained using Method III and
  Eq.~\eqref{eq:1Dform},
  we see that the direction-averaged 1D form of the Yaglom law, which 
  \emph{assumes} inertial range conditions,
  actually produces a smaller estimate for the cascade rate, with a peak 
  of $\sim 83\%$ of $\epsDiss$.

\begin{figure}[ht]
    \centering
    \includegraphics[width = 0.49\textwidth]{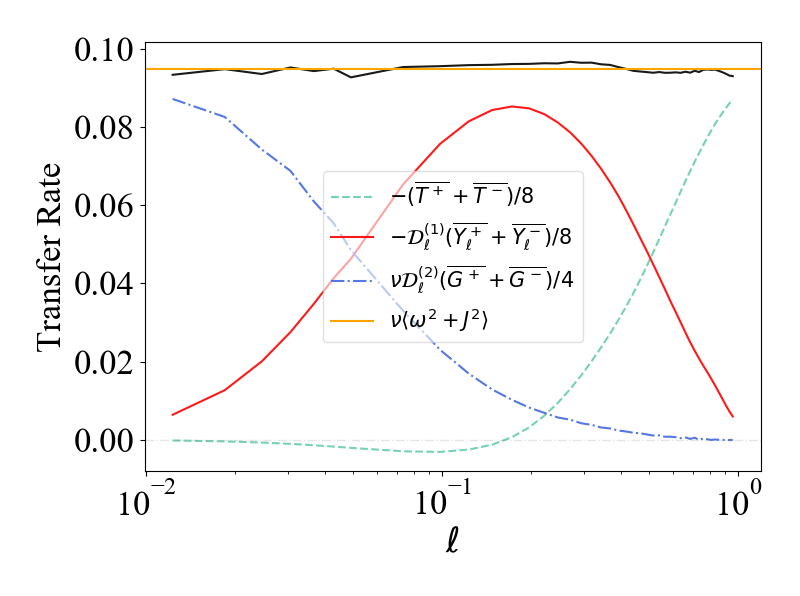}
    \caption{Terms in the 
    direction-averaged  
    \vKH equations, same as Fig.~\ref{fig:ygt-all-I}, except using data from anisotropic simulation III ($B_0 = 2$, high $\sigma_c $). Energy balance is obtained 
    at all scales, as indicated by the comparison of 
    the actual dissipation rate calculated from $\nu \langle \omega^2 + J^2 \rangle$
    with the black line, which is the sum of the three lines representing the $Y$, $G$ and $T$ terms. 
    }
    \label{fig:lsb2-ygt-III}
\end{figure}

The {equivalent} results for the decaying anisotropic simulation III with global field $\vct{B}_0 = 2\uv{z}$ are shown in Fig.~\ref{fig:lsb2-ygt-III}. 
{Energy balance is again evident, in this case at all scales.}
  {Furthermore, 
  the peak value of the 
 ${-\cal D}^{(1)}_\ell(\overline{Y^+_\ell}+\overline{Y^-_\ell})/8$
 curve 
  is
 $\sim 87 \% $ of $\epsDiss$, 
 which is usefully close to the true value.}

\subsection{Comparison between anisotropic cases} \label{subsec:gresults}

We recall that two anisotropic simulations (runs II and III) are included, in part to explore the parameter variations that may influence the results.
These simulations have different magnetic field strengths and different levels of cross helicity, both of which are known to influence the nature of the anisotropic cascade \citep{PouquetEA86,PolitanoPouquet98-vKH,OughtonEA15}.

\begin{figure}[ht]
    \centering
    \includegraphics[width = 0.49\textwidth]{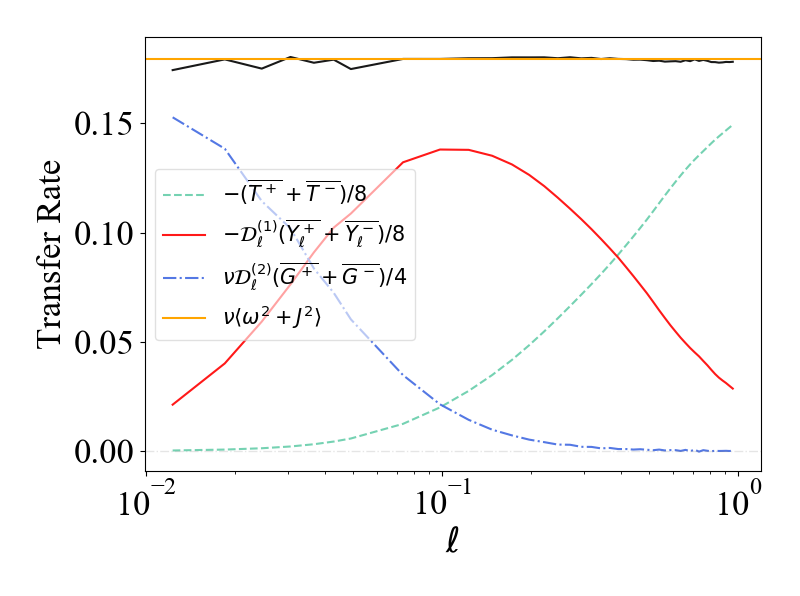}
    \caption{Terms in the 
    direction-averaged  
    \vKH equations, same as Fig.~\ref{fig:lsb2-ygt-III}, except using data from anisotropic simulation II ($B_0 = 1$, low $ \sigma_c$). Energy balance is obtained 
    at all scales.
    }
    \label{fig:nsl-ygt-II}
\end{figure}

In this section we make a comparison between the results from these two anisotropic cases, simulations II and III. First, as a comparison with Fig.~\ref{fig:lsb2-ygt-III} for simulation III, we show in Fig.~\ref{fig:nsl-ygt-II} terms of the direction-averaged \vKH equation (Eq.~\eqref{eq:avereduced}) for simulation II, which add up to the exact dissipation rate. 
However, for simulation II, due to the non-negligible effect of the time variation and dissipative terms, we observe that the peak value of the ${-\cal D}^{(1)}_\ell(\overline{Y^+_\ell}+\overline{Y^-_\ell})/8$ curve is much smaller than the exact dissipation rate (approximately $77\%$). 
Thus, one can also expect that using the direction-averaged 
1D form third-order law (Method III), which only considers the Yaglom term $\overline{Y^{\pm}_\ell}$, 
will yield a less accurate estimate for the cascade rate.

As discussed in section~\ref{sec:background}, large Reynolds numbers are required in order to have an inertial range that is well-separated from the dissipation range. 
Unfortunately, due to limited computing capabilities, 
this is not the case for our simulations.  
Furthermore, it has been assumed that the time variation term is negligible in Eq.~\eqref{eq:fullform3d}, along with the dissipation term. Again, this is not the case for the anisotropic simulations we report on herein (see discussion of Figs.~\ref{fig:lsb2-ygt-III} and \ref{fig:nsl-ygt-II}).
Since our simulation II is not in a regime where the 
cascade  terms (\divYpm)
dominate over the time variation and dissipation terms, the  transfer rates estimated using a third-order law are significantly less than the actual dissipation rate, even in lag directions perpendicular to $\vct{B}_0$,  which usually 
give values closer to the exact energy transfer rate
(see Fig.~\ref{fig:nsb1-1d}).
Of the runs we consider simulation III is perhaps the least limited in this respect. 
In particular, as Fig.~\ref{fig:lsb2-ygt-III} indicates,
the \textbf{peak}
${-\cal D}^{(1)}_\ell(\overline{Y^+_\ell}+\overline{Y^-_\ell})/8$
 contribution is $\sim 13\%$ smaller than $\epsDiss$.
  {However, the situation is complicated since}
in some parts of the inertial range there is cancellation of the (negative at small $\ell$) time variation term $-(\overline{T^+}+\overline{T^-})/8$ 
and the dissipative term $\nu {\cal D}^{(2)}_\ell (\overline{G^+}+\overline{G^-})/4$.

\section{Cascade rate and single spacecraft sampling}\label{sec:spacecraft}

Single spacecraft observations in the solar wind provide lags in only one direction, thus we can apply only the 1D form third-order law, 
usually that written as Eq.~\eqref{eq:ez}; 
and described as Method I, see \S\ref{subsec:theory}. 
The same definition $Y^\pm_\ell = \langle (\uv{\ell} \cdot \delta \vz^\mp) |\delta \vz^\pm|^2 \rangle$ is employed,  but now 
$\vz^{\pm} = \vct{v} \pm \vct{B}/ \sqrt{\mu_0 \langle n_i \rangle m_i }$, where $\mu_0$ is the magnetic permeability of free space and $m_i, \langle n_i \rangle$ are respectively the mass and (interval averaged) number density of the solar wind protons.
Here we analyze measurements made by Parker Solar Probe from Nov.\ 3--8, 2018;
the time cadence of the data is 1\,s.  The purpose is not to provide an exhaustive
treatment of the solar wind cascade rates. Instead we present 
an example to inform and support our discussion. 

\begin{figure}[ht]
    \centering
    \includegraphics[width = 0.5\textwidth]{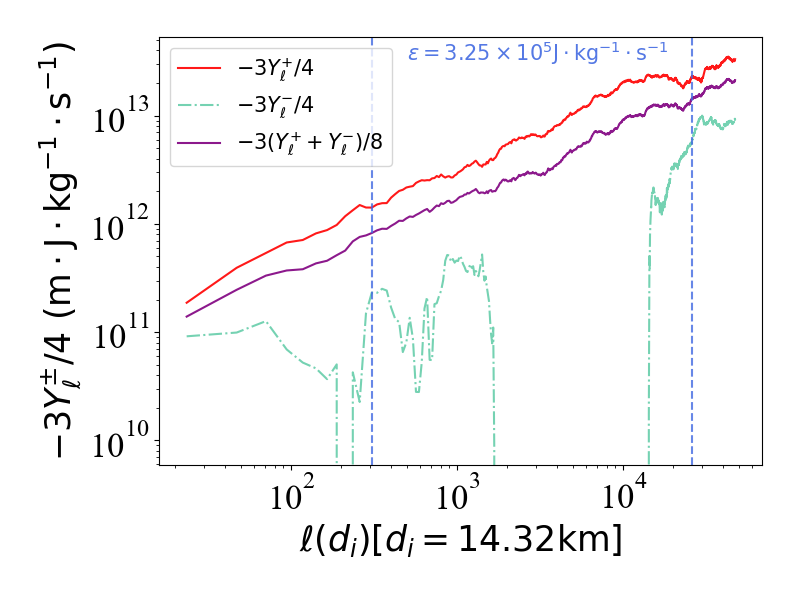}
    \caption{The 1D form third-order law {(Method I)} is used for Parker Solar Probe observations to estimate the energy dissipation rate (data from Nov.\ 3--8 2018, with averaged helio-distance 37.9 solar radii; lags are in units of the ion inertial length, $d_i$). 
    The indicated value of $\epsilon =  3.25 \times 10^5$ J/kg-s is obtained using a
    least mean square linear fit to $ -3(Y^+_\ell+Y^-_\ell)/8 = \epsilon \ell$. 
    The inertial range in which we 
    perform the fit to the 1D form is indicated by the two vertical dashed lines.}
    \label{fig:psp-data}
\end{figure}

The dissipation rate $\epsDiss$ can be estimated by performing a linear fit in the inertial range, where the corresponding slope gives the value of $\epsDiss$.
We choose the inertial range as the range separated from the correlation length and the estimated scale at which kinetic effects become important. The latter is typically a few ion inertial scales, as indicated 
in Fig.~\ref{fig:psp-data}.
Generally we also examine the energy spectra (not shown) 
to ensure that a reasonable power-law distribution is found
in the selected range of scales. A linear least mean square method is employed over the selected inertial range for determining the best fitting slope to the computed values of $Y_\ell^\pm$. 
Note that Fig.~\ref{fig:psp-data} is plotted in log-log form, while the
best fits are computed in linear-linear space. 
One should not be confused with the slope of the line in log-log form and the 
estimated dissipation rate $\epsilon$. Plotted this way (log-log) the slope of the line indicates the power law in lag, which is expected to be linear, while the level of the line determines the estimated dissipation rate.

Third-order statistics are notoriously noisy in the solar wind
and a single example will not be fully representative. 
The sample result shown can be meaningfully 
compared with other recent third-order analyses, including those
 that employ PSP data \citep{andres2021evolution,andres2022incompressible,BandyopadhyayEA20-PSP3rd}. However, detailed comparisons are beyond the scope of the current study. 
 Recent studies also indicate correlation anisotropy \citep{Cuesta_2022} and  anisotropy of the energy spectrum  \citep{huang2022anisotropy,zhang2022three} in the PSP datasets, which may suggest anisotropy of nonlinear energy transfer. More evidence has been shown in \cite{andres2022incompressible}, which examines variations of third-order law with decomposition into parallel and perpendicular components using PSP data.

As discussed earlier, in anisotropic configurations, the dissipation rate we obtain from the 1D form third-order law Eq.~\eqref{eq:ez} depends on the lag direction, as is evident in Fig.~\ref{fig:lsb2-1d-dd} for simulation III and in Fig.~\ref{fig:nsb1-1d} for simulation II. In addition, the inertial range is also not uniquely determined in anisotropic cases and estimates of the optimal range of lags to associate with a putative inertial range vary with the lag direction as seen in the same figures. 
Therefore, the estimates of dissipation rate vary when analyzing different directions and assuming different inertial ranges. 
Nonetheless, the 1D form third-order law 
may provide 
reasonable approximations to the 
actual dissipation rate if the 
directional variability in 
values of estimated dissipation rate 
are acceptable. 
For example, cascade rates estimated from different peaks in Fig.~\ref{fig:lsb2-1d-dd} vary by about 50\% of the exact dissipation rate.

The solar wind has a global magnetic field, and, in the inner heliosphere, a relatively high cross helicity. These features are similar to those of simulation III. Examining the results of our analysis for this run (Fig.~\ref{fig:lsb2-1d-dd}, right panel), the error in the estimates of the cascade rate obtained from the 1D form third-order law can usually be assessed from the variation in the peak value and  the variation in the location of the peak in the lag axis.

\section{Discussion and Conclusions}\label{sec:conc}

We have examined the properties of several formulations
for analyzing energy transfer in homogeneous MHD turbulence.
The \vKH equations 
in increment form, Eq.~\eqref{eq:fullform3d},
symbolically written as Eq.~\eqref{eq:fullform-sym}, provide the most complete treatment. These are exact equations and account for dissipation,
time dependence, 
{nonlinear transfer,} 
and anisotropy. 
Quantitative evaluation of the 
several terms in the \vKH equations
affords direct insight into the conditions 
required to identify a range of scales that can reasonably be viewed as an \emph{inertial range}. 
Ideally in such a range, Kolmogorov's assumptions of steady dissipation-free transfer across scales
can be realized, and the only term that makes substantial contribution is the one involving the third-order structure functions, $\vY^\pm$. In that case various forms of the Kolmogorov--Yaglom \citep{Frisch}
law become relevant,
and specifically in MHD, the third-order law 
derived first by 
\cite{PolitanoPouquet98-grl}. 

When less information is available, and in particular when it is impractical to 
determine three-dimensional derivatives in lag space,
researchers have traditionally adopted one of several approaches to simplify the estimation of the transfer rate, which in steady conditions is the cascade (or dissipation) rate. We have examined several issues that affect these familiar approximations.

In our study, we examined both the \vKH equation and a simpler form of the third-order law for incompressible MHD simulations of turbulence systems. 
In MHD under different global magnetic field conditions, one can compute the correct value for the cascade rate if each term of the \vKH equation, Eq.~\eqref{eq:fullform3d} or \eqref{eq:fullform-sym}, 
can be computed exactly.
{With a numerical discretization only approximate values can be obtained.}
Nevertheless, averaging the \vKH equation in a sufficient number of independent directions (e.g., spanning a spherical surface),
and keeping only the radial ($\uv{\ell}$) components of the so-obtained $\vY^\pm$, 
can provide accurate results.
This directional-averaging strategy is based on a rigorous reduction of the problem to a one dimensional direction-averaged form, 
a direct extension to MHD of the hydrodynamic result due to \citet{NieTanveer99}.
The direction-averaging approach 
may be particularly useful when 
adapted for use with multi-spacecraft datasets (for which there are typically only a small number of lag directions available) to estimate local cascade rates for space plasma turbulence. 
The accuracy of this method for different simulations is reported on in Section~\ref{sec:results_3Dlaw} and displayed visually in Figures~\ref{fig:ygt-all-I}, \ref{fig:lsb2-ygt-III}, and \ref{fig:nsl-ygt-II}.

For further simplification, we require the existence of an inertial range of substantial length (i.e., very high Reynolds number) to justify the use of what we have called the third-order laws (\S\ref{sec:third-order}). 
Since the condition of infinite Reynolds number cannot be achieved, we are not able to observe a perfect inertial range, thereby leading to discrepancies between the actual cascade rate and the one determined from the third-order term. 
An advantage of these simplified forms is 
that they ignore
dissipative and time variation effects, which are difficult to measure experimentally. 
The 3D form of the third-order law requires that at least some measurements are available in different directions and gives a reasonable estimate of 3D derivatives in lag space in simulations \citep{VerdiniEA15}. 
In the presence of time stationarity, the 3D form third-order law can provide an accurate estimate of energy transfer rate in the inertial range.

The futher assumption of isotropy is often used for in situ measurements of solar wind turbulence \citep[e.g.,][]{StawarzEA09, OsmanEA11-3rd} and magnetic reconnection in Earth's magnetosheath \citep[e.g.,][]{BandyopadhyayEA21-mtail}. 
In the presence of isotropy, the 3D form third-order law can be simplified to a 1D form, which employs only one lag direction.
For isotropic MHD simulations, the 1D form third-order law provides a reasonable approximation, with some statistical variation 
with changing lag direction. 

Issues regarding
accuracy of energy transfer 
rate estimation become still more significant  
when anisotropy is induced by a mean magnetic field, a 
circumstance expected to be of significance in space and astrophysical 
plasmas.   
We report on two anisotropic simulations (Figs.~\ref{fig:lsb2-1d-dd} and \ref{fig:nsb1-1d}) in Section~\ref{subsec:1D_comps} and (Figs.~\ref{fig:lsb2-ygt-III} and \ref{fig:nsl-ygt-II}) in Section~\ref{subsec:gresults}.
From these results, 
it is apparent that 
error in estimation from the 
1D form third-order law follows from a combination of effects due to the magnitude of the global magnetic field (anisotropy), the contribution of the dissipative term, and a lack of time stationarity.
These errors can add up to be $\sim 50\%$ of the exact dissipation rate in our simulations.  We conclude that the (unaveraged) 1D form provides a correct order of magnitude 
result for the moderate levels of anisotropy found in the parameters we adopted. 

\begin{figure}[ht]
    \centering
    \includegraphics[width = 0.6\textwidth]{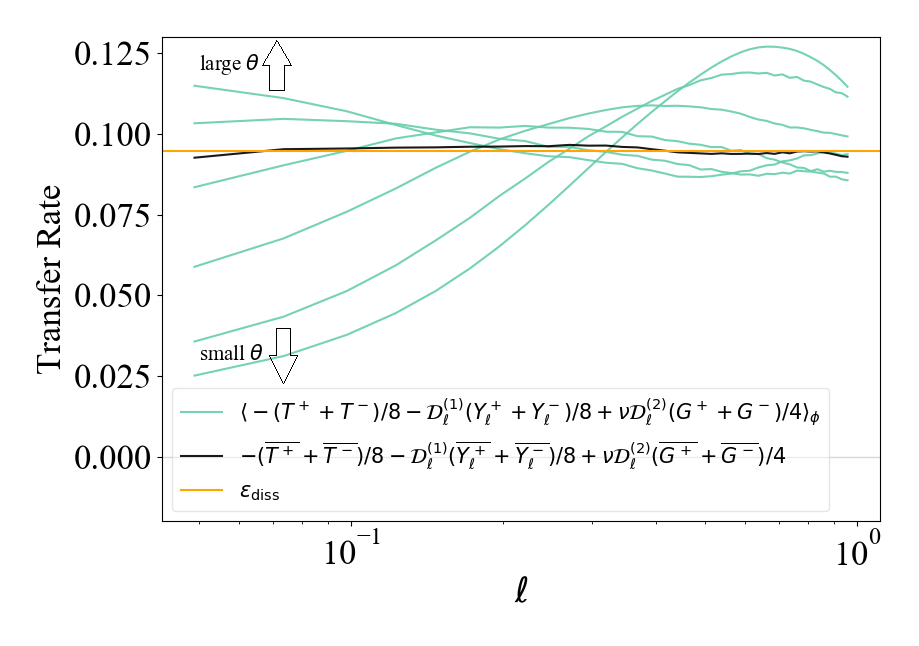}
    \caption{
    Two sorts of angular averaging of the \vKH equation Eq.~\eqref{eq:vKH-epsDiss} applied to data from the $B_0 = 2$, high $\sigma_c$ simulation III.
    Terms with an overbar 
    indicate an average over full $4\pi$ solid angle (all lag directions distributed uniformly on a sphere); see Eq.~\eqref{eq:avereduced}. 
    Each green curve 
    is calculated for a different fixed polar angle $\theta$ averaged over 12 azimuthal angles $\phi$, with the averaging denoted by $\langle -\rangle_{\phi}$.
    The black curve is the averaged value of all the green curves after weighting by $\sin{\theta}$.
    }
    \label{fig:ygtsub}
\end{figure}

Several aspects of
the effects of anisotropy are  summarized in Fig.~\ref{fig:ygtsub}, which includes an assessment of the 
direction-averaged 
\vKH equation Eq.~\eqref{eq:avereduced}.
We intend to highlight variations 
of energy transfer for different polar angles $\theta$,
and therefore 
we do not show variations with 
varying azimuthal angle $\phi$.
In Fig.~\ref{fig:ygtsub}, we plot (green curves)
the sum of the time variation ($-(T^+ + T^-)/8$), 
the 1D Yaglom term (${-\cal D}^{(1)}_\ell (Y_\ell^+ + Y_\ell^-)/8$),
and the 1D dissipative term (${\nu \cal D}^{(2)}_\ell (G^+ + G^-)/4$), each averaged over 12 azimuthal angles for fixed polar angles ranging over the interval $0\le \theta \le 90^{\circ}$.
Recall the definitions of the 1D operators ${\cal D}^{(1)}$ and 
${\cal D}^{(2)}$ are given below Eq.~\eqref{eq:avereduced}.
The average over $\phi$ is indicated by 
$\langle \dots \rangle_\phi$ in the legend. 
Note that the procedure is not equivalent 
to 
full $4 \pi$ solid angle averaging, 
which is indicated by an overbar 
as in Eq.~\eqref{eq:avereduced}.
Clearly averaging over $\phi$ is only a partial averaging
and, in general, falls short of the effect of fully directional averaging. 

The 
average of the contributions from each $\theta$ 
can be weighted by $\sin{\theta}$  and summed, to provide 
an approximation to a proper average over the sphere, which is the full $4 \pi$ solid angle averaging 
indicated by the overbar 
as in Eq.~\eqref{eq:avereduced}. 
This sum is also shown in Fig.~\ref{fig:ygtsub} (as the black line). We observe that this summation 
closely adds up to the exact dissipation rate, also shown 
(orange line). This demonstrates the approximate convergence 
that is expected based on the MHD generalization of 
the \cite{NieTanveer99} exact result for hydrodynamnics,
as we discussed in previous sections.

To delve somewhat further into this analysis of anisotropy, we note that
since each partially averaged (green) curve in Fig.~\ref{fig:ygtsub} is obtained 
at one only polar angle $\theta$ relative to the mean field, 
we see a large variance of the sum of the three (LHS) terms in the \vKH equation. 
At small scales, the top curve, having the largest 
estimated total transfer rate, corresponds to $\theta (\vct{\ell}, \vct{B}_0)$
close to $90^{\circ}$ and the bottom curve, with lowest estimated transfer rate, 
corresponds to $\theta (\vct{\ell}, \vct{B}_0)$ near $0^\circ$, almost parallel to the mean field. 
In fact, the estimated transfer rate at small scales increases monotonically 
as we increase the polar angle (from $0^\circ$ to $90^{\circ}$).
On the contrary, at large scales, the sum of the three terms is larger for smaller $\theta$.
As a consequence, the peaks of these curves shift from large scales to small scales as we increase $\theta$.
In addition, we may notice that these curves are intertwined in the middle range of lags, which is approximately the inertial range  
  (see Fig.~\ref{fig:lsb2-ygt-III}). There is actually a rather narrow range of lags near
  $\ell = 0.3$ in which the estimates derived from different polar angles are close to one 
  another, varying by only about $\pm 10\%$.
Overall, the observed 
variability at different values of polar angle $\theta$ 
emphasizes the necessity of a uniform angular coverage 
of the lag directions to compute an accurate energy transfer 
rate in anisotropic cases. 
{It is also consistent with} variability associated with the 
$ Y^\pm_\ell / \ell $ term 
 (i.e., Method I)
that was demonstrated for the anisotropic cases 
  (Figs.~\ref{fig:lsb2-1d-dd} and \ref{fig:nsb1-1d}), 
and to a lesser extent 
even for the isotropic case (Fig.~\ref{fig:ssb0-1d-dd}).


This paper has been developed with two main intentions.  First it is intended to 
summarize pertinent 
analytical results relating
to measurement of energy transfer rates 
beginning with the \vKH equations 
and leading to several 
reductions that are essentially 
third-order (Yaglom) laws,  
or for MHD, Politano--Pouquet third-order laws.
The reduction to Yaglom laws becomes
applicable when an inertial range is present, so  that all terms but
the Yaglom flux become negligible in contributing to the total energy balance 
over a range of lags in the inertial range.  
The second purpose has been to provide examples 
and caveats concerning 
the use of these methods by application to several moderate 
resolution MHD turbulence simulations. 

The results are generally seen to be encouraging.
Even with significant variation 
in estimates expected due to variation of lag direction in anisotropic cases, and due to ``pollution'' of the putative inertial range by time-dependent 
and dissipative effects, estimates can be broadly accurate within tolerances of $\sim $50$\%$. 
Simulations can generally do better than this, but for single spacecraft observations this may 
be an acceptable estimate. 
It is clear that estimates will improve 
when three-dimensional derivatives in lag space are available, and when a large number of baseline directions are available so one might approach optimal direction-averaging.  
Some progress has been realized along these lines by exploiting these features in the four spacecraft \textit{Cluster} mission \citep{OsmanEA11-3rd}.
Significant advances in 
evaluating inertial range transfer rates will become available in the 
Helioswarm mission, comprising 9 spacecraft and currently under development \citep{SpenceEA19,KleinEA19,MatthaeusEA19} and a larger 24 point 
configuration envisioned in the 
MagneToRE approach \citep{maruca2021magnetore}.

Finally we mention some limitations of the 
present study. We have not 
attempted any examination of the accuracy of third-order laws 
in differing interplanetary conditions and therefore 
have shown only a single case of solar wind cascade rate analysis. 
A complete study of solar wind situations would inevitably require examining
variations of a number of interplanetary parameters 
including fluctuation and mean field strength, wind speed, cross helicity,
turbulence age, etc. Such an effort is highly worthwhile, 
and the present study provides
some guidelines regarding how such a major study might be undertaken, 
but for brevity and focus, we
defer attempts at a comprehensive comparison 
with varying solar wind conditions to future research.

The present study is also limited to incompressible and simple
MHD cases. Including compressibility,  Hall effects and additional physical influences on energy transfer introduces considerable additional complexity to the estimation of the inertial range transfer rate, and requires much more extensive discussion. Some relevant observational results include some of these more general physical descriptions. For example, recent studies employing Parker Solar Probe, MAVEN, Cluster, Magnetosphere Multiscale and THEMIS data 
 \citep{banerjee2016scaling,hadid2017energy,HadidEA18,AndresEA19,andres2021evolution} found moderate increases of compressible energy transfer rate with respect to the incompressible transfer rate.  These results point the way to future studies that would generalize the simpler case that we have examined here.

\begin{acknowledgments}
This research supported in part by
the NASA Parker Solar Probe Mission
under a GI grant 80NSSC21K1765 and the 
ISOIS team (Princeton SUB0000165),
by the IMAP project (Princeton
SUB0000317), by the MMS mission under a 
Theory and Modeling grant  80NSSC19K0565,
by NASA HSR grants
80NSSC18K1648 and 80NSSC19K0284,
and by the US National Science Foundation NSFDOE program 
under grant PHY2108834. 
\end{acknowledgments}

%






\appendix

\section{Reduction of 3D to 1D forms by directional averaging} 
\label{appA}

Starting from Eq.~\eqref{eq:fullform3d}, let us integrate it over a spherical surface (i.e., over solid angle),
using spherical polar coordinates with $\vct{\ell}$ the ``radius.'' See \cite{NieTanveer99}
and \cite{TaylorEA03}
for the closely related hydrodynamic case.
As in Eq.~\eqref{eq:fullform-sym}, we adopt abbreviations: 
$\frac{\partial}{\partial t} \langle (\delta \vz^{\pm})^2 \rangle = T^{\pm}$, 
$\langle \delta \vz^{\mp} |\delta \vz^{\pm}|^2 \rangle = \vY^{\pm}$, 
$\langle (\delta \vz^{\pm})^2 \rangle = G^{\pm}$. 

The $\vY^{\pm}$ term of Eq.~\eqref{eq:fullform-sym} becomes
\begin{equation}
\begin{split}
    \int_S \grad_{\ell} \cdot \vY^{\pm}\d\Omega &= \int_0^{\pi}\int_0^{2\pi} \left[ \frac{1}{\ell^2} \frac{\partial}{\partial \ell}(\ell^2 Y_{\ell}^{\pm}) + \frac{1}{\ell\sin{\theta}}\frac{\partial}{\partial \theta}(\sin{\theta} Y_{\theta}^{\pm}) + \frac{1}{\ell\sin{\theta}}\frac{\partial}{\partial \phi} Y_{\phi}^{\pm} \right] \, \sin{\theta} \d\phi \d\theta \\
     &= \int_0^{\pi}\int_0^{2\pi}\frac{1}{\ell^2} \frac{\partial}{\partial \ell}(\ell^2 Y_{\ell}^{\pm}) \sin{\theta} \, \d\phi \d\theta + \int_0^{2\pi}\frac{1}{\ell}(\sin{\theta}Y_{\theta}^{\pm})\bigg|_0^{\pi} \, \d\phi + \int_0^{\pi}\frac{1}{\ell}(Y_{\phi}^{\pm})\bigg|_0^{2\pi} \,\d\theta. \\
\end{split}    
\end{equation}
By periodicity, the integrals of the  $\theta$ and $\phi$ components vanish and therefore
\begin{equation}
    \int_S \grad_{\ell} \cdot \vY^{\pm} \d\Omega =  \int_0^{\pi}\int_0^{2\pi} \frac{1}{\ell^2} \frac{\partial}{\partial \ell}(\ell^2 Y_{\ell}^{\pm}) \sin{\theta} \d\phi \d\theta.
\label{y_intermed}
\end{equation}
Similarly, writing the Laplacian of $G^{\pm}$ term in spherical polar coordinates also, and then integrating over a spherical surface,
\begin{equation}
    \begin{split}
    \int_S 2\nu \nabla^2_l G^{\pm} \d\Omega &= \int_0^{\pi} \int_0^{2\pi}  \frac{2\nu}{\ell^2} \left[\frac{\partial}{\partial \ell}\left(\ell^2 \frac{\partial G^{\pm}}{\partial \ell}\right) + \frac{1}{\sin{\theta}} \frac{\partial}{\partial \theta}\left(\sin{\theta}\frac{\partial G^{\pm}}{\partial \theta}\right) + \frac{1}{\sin^2{\theta}} \frac{\partial^2 G^{\pm}}{\partial \phi^2} \right] \sin{\theta} \d\phi \d\theta 
    \\
    &=
    \int_0^{\pi}\int_0^{2\pi}\frac{2\nu}{\ell^2} \frac{\partial}{\partial \ell}\left(\ell^2 \frac{\partial G^{\pm}}{\partial \ell}\right)\sin{\theta} \d\phi \d\theta
    \end{split}
\end{equation}
where the $\partial_\theta$ and $\partial_\phi$ dependent terms also vanish after integration; in a numerical (discretized) evaluation this vanishing relies on a proper distribution of lag directions.

Using these results we can write the integral of Eq.~\eqref{eq:fullform-sym} over {solid angle} 
as
\begin{equation}
\int_0^{\pi}\int_0^{2\pi} T^{\pm} \sin{\theta} \,\d\phi \d\theta + \int_0^{\pi}\int_0^{2\pi} \frac{1}{\ell^2} \frac{\partial}{\partial \ell}(\ell^2 Y_{\ell}^{\pm}) \sin{\theta} \d\phi \d\theta
- \int_0^{\pi}\int_0^{2\pi}\frac{2\nu}{\ell^2} \frac{\partial}{\partial \ell}\left(\ell^2 \frac{\partial G^{\pm}}{\partial \ell}\right)\sin{\theta} \d\phi \d\theta
= - 16\pi \epsilon^{\pm}, 
\label{num_fullygt}
\end{equation}
where 
$4\epsilon^{\pm} \int_0^{\pi}\int_0^{2\pi} \sin{\theta}\, \d\phi \d\theta = 16\pi \epsilon^{\pm}$ gives the last term. 
Abbreviating the average over full $4\pi$ solid angle by an overbar, this equation leads immediately to
Eq.~\eqref{eq:avereduced}.

When analyzing simulation data, these integrals are evaluated using discrete approximations. 
To illustrate this we consider the simpler well-defined inertial range case, the 3D form of the third-order law, Eq.~\eqref{eq:3d_ez}, integrated over solid angle.
Using Eq.~\eqref{y_intermed} in Eq.~\eqref{eq:3d_ez}, multiplying by $\ell^2$, and integrating over both $\d\Omega$ and $\ell$ yields
\begin{equation}
\int_0^{\pi}\int_0^{2\pi}Y_\ell^{\pm} \sin{\theta} \, \d\phi \d\theta = -\frac{4}{3} \epsilon^{\pm} \ell  \int_0^{\pi}\int_0^{2\pi} \sin{\theta} \, \d\phi \d\theta.    
\label{3d_y_epm}
\end{equation}
If the average over full $4\pi$ solid angle is again denoted by an overbar, 
this result states that 
\begin{equation}
\overline{Y_{\ell}^{\pm}} = 
\overline{\langle (\uv{\ell} \cdot \delta \vz^\mp) |\delta \vz^\pm|^2 \rangle}
= -\frac{4}{3} \epsilon^{\pm} \ell .
\label{thirdorderaveraged}
\end{equation}
This is written as 
Eq.~\eqref{eq:1Dform} in the main text.
With a simple discretization of $\theta$ and $\phi$ this becomes 
\begin{equation}
    -\frac{3}{4} \frac{\sum_i \sum_j Y_{\ell}^{\pm}(\theta_i, \phi_j) \sin{\theta_i} \, \Delta \theta_i \Delta \phi_j}{4\pi} = \epsilon^{\pm} \ell .
    \label{discret}
\end{equation}
Similar discretization approaches are applied to the more general case, the direction-averaged \vKH equation, Eq.~\eqref{eq:avereduced} or Eq.~\eqref{num_fullygt}.

 \newcommand{\BIBand} {and} 
  \newcommand{\boldVol}[1] {\textbf{#1}} 
  \providecommand{\SortNoop}[1]{} 
  \providecommand{\sortnoop}[1]{} 
  \newcommand{\stereo} {\emph{{S}{T}{E}{R}{E}{O}}} 
  \newcommand{\alfven} {{A}lfv{\'e}n\ } 
  \newcommand{\alfvenic} {{A}lfv{\'e}nic\ } 
  \newcommand{\Alfven} {{A}lfv{\'e}n\ } 
  \newcommand{\Alfvenic} {{A}lfv{\'e}nic\ }



\end{document}